\documentclass[12pt,preprint]{aastex}

\shorttitle{Relationship between Hard and Soft X-rays}
\shortauthors{Guo et al.}

\begin{document}

\title{Relationship between Hard and Soft X-ray Emission Components
of a Solar Flare}

\author{Jingnan Guo \altaffilmark{1,2,3}, Siming Liu \altaffilmark{1,3},
Lyndsay Fletcher \altaffilmark{1}, and Eduard P. Kontar
\altaffilmark{1}}

\altaffiltext{1}{School of Physics and Astronomy, SUPA, University of
Glasgow, Glasgow, G12 8QQ, Scotland; j.guo@astro.gla.ac.uk}
\altaffiltext{2}{Max-Planck Institute for Solar System Research,
Katlenburg-Lindau, 37191, Germany} \altaffiltext{3}{Purple
MountainObservatory, Nanjing 210008, China}

\begin{abstract}

X-ray observations of solar flares routinely reveal an impulsive
high-energy and a gradual low-energy emission component, whose
relationship is one of the key issues of solar flare study. The
gradual and impulsive emission components are believed to be
associated with, respectively, the thermal and nonthermal components
identified in spectral fitting. In this paper, %an example of
a prominent $\sim 50$ second hard X-ray (HXR) pulse of a simple GOES class C7.5 flare on 20 February 2002 is used to study the association between high energy, non-thermal and impulsive evolution, and low energy, thermal and gradual evolution.
We use regularized methods to obtain time derivatives of
photon fluxes to quantify the time evolution as a function of photon
energy, obtaining a break energy between impulsive and gradual
behavior. These break energies are consistent with a constant value of $\sim 11$ keV in agreement with those found spectroscopically between thermal and non-thermal components, but
the relative errors of the former are greater than $15\%$ and much greater than
the a few percent errors found from the spectral fitting. These errors only weakly depend on assuming an
underlying spectral model for the photons, pointing to the
current data being inadequate to reduce the uncertainties rather
than there being a problem associated with an assumed model.
The time derivative method is used to test for the presence
of a `pivot energy' in this flare. Although these pivot energies are marginally consistent with a constant value of $\sim 9$ keV, its values in the HXR rise phase appear to be lower than those in the decay phase.
Assuming that electrons producing the high-energy component have a power law distribution and are accelerated from relatively hot regions of a background plasma responsible for the observed thermal component, a low limit is obtained for the low-energy cutoff. This limit is always lower than the break and pivot energies and locates in the tail of the Maxwellian distribution of the thermal component.

\end{abstract}


\section{INTRODUCTION}
\label{intro}

High-energy observations of solar flares with the Reuven Ramaty
High-Energy Solar Spectroscopic Imager (RHESSI)
\citep{lin2002rhessi} allows high resolution studies over a broad energy range from 3 keV soft X-rays to $\gamma$-rays up to 17 MeV.  The
photon flux in the energy range of $\sim 20 - 100$~keV can be reasonably well fitted with a
power-law function, and its time-variability increases with the
photon energy \citep{asch2005psc,mcateer2007bursty}. It is commonly assumed that
this emission is produced by an electron population distinct from
electrons forming a thermal background plasma, which is presumed to
produce the low-energy X-ray emission \citep[e.g.][]{asch2002paa}. The impulsive high-energy emission originates predominantly from the
chromospheric footpoints, while --- at least later in the flare ---
the more slowly-varying low-energy emission is dominated by a hot
coronal source, observed in many cases to be located near EUV flare
loops \citep[e.g.][]{gallagher2002}. These observations are
usually interpreted in the framework of the standard flare model
where the hard X-ray (HXR) emission at the chromospheric footpoints
of magnetic loops is bremsstrahlung of non-thermal high-energy
electrons moving downward along flare loops from acceleration sites
higher up in the corona \citep{brown1971}, with the resulting
footpoint heating and evaporation leading to the hot (usually dense)
coronal thermal component  \citep{neupert1968, petrosian1973,
fisher1989}. Note, we do not automatically adopt this assumed relationship between accelerated and heated particles. In fact, in Section~\ref{s:discussion}, we interpret the observations in a framework where the non-thermal electrons are accelerated out of a heated thermal background.
The `non-thermal' electron distribution is usually
assumed to have a low-energy cutoff, the presence of which ensures
that the total electron number and power are finite. However, it is
not clear that there is a theoretical mechanism for particle
acceleration which can naturally lead to the low-energy cutoff
distinguishing non-thermal from thermal particles \citep{benz1977,
miller1997jgr, petrosian_liu2004}. Indeed, it has been argued by
\citet{emslie2003} that the low energy cutoff may be a
redundant concept. \citet{hannah2009wave} suggested that a sharp cutoff in the injected electron spectrum disappears with the inclusion of wave-particle interactions. A dip in the electron distribution obtained through the inversion of the observed photon spectrum of some flares may be associated with the low-energy cutoff. \citet{Kontar2008cutoff}, however, showed that such a feature vanishes when isotropic albedo correction is applied.

The time correlation between the impulsive HXR and/or radio
emission and the derivative of the gradual emissions at certain
energies, the so-called `Neupert effect',  \citep{neupert1968,
dennis1993} carries with it the implication that most gradual
emissions are a ``by-product", resulting from energy deposition by
non-thermal electrons. This is also suggested by the high
non-thermal electron energy content resulting from application of
the standard collisional thick-target model
\citep[e.g.][]{emslie2004,emslie2005}, which points to
high-energy electrons forming a dominant channel in the energy
conversion process. However, more quantitative examination of relevant observations show that the picture is somewhat less clear. Some flares involve heating of thermal coronal plasma in
the absence of a power-law emission component \citep{battaglia2009},
many show footpoints with impulsive phase emission within the energy
range usually considered as thermal \citep{mrozek2004}, and
the `Neupert effect', which never represents a perfect correlation,
does not hold in all flares or at all (thermal) energies
\citep{mctiernan1999,veronig2002, veronig2005}. So the possibility
of heating of the solar plasma as a direct part of the energy
release before and/or during the acceleration is still an open issue
for investigation \citep{petrosian_liu2004, liu2009elementary}. By deriving abundances of elements with low first ionization potentials, such as calcium and iron, \citet{feldman2004} found that at least the hot plasmas of some flares result from direct in situ heating of corona plasma, possibly due to a compression process. This approach may also lead to a measurement of the partition of hot flare plasmas originated from the corona and chromosphere.

The general question of how the pre-flare magnetic energy is
converted into radiation, plasma bulk motion, thermal and
non-thermal particle energy may not have a simple answer
\citep{emslie2005}. Although flares share the same kind of energy
source, different flares can have quite different appearances, and
possibly involve different physical processes. Nevertheless, some
well-observed characteristics can still set constraints on the
overall energy dissipation process. The soft-hard-soft spectral
evolution of some HXR pulses is one of the most important
characteristics of high-energy emissions
\citep{kane1970,grigis2004spectral} and may point to a
turbulent particle acceleration mechanism \citep{grigis05}. Early
analyses \citep{gan1998invariable} suggested that there is a value
of photon energy at which the non-thermal flux does not change,
so that the power-law  pivots about this location, a possible further model constraint. \citet{grigis2004spectral} showed that there is no single `pivot energy', rather there is a small range.  \cite{battaglia2006} determined that in the rise phase this energy may be lower than that in the decay phase.  In the context of stochastic particle acceleration from the thermal  background plasma,  the pivot energy should evolve with the background plasma properties
\citep{petrosian_liu2004,liu2010elementary}. However,
distinguishing between different models on the basis of observations
remains a challenging task \citep{grigis05}.

Other constraints based on the evolution of HXR light curves include
the observation that sub-second HXR pulses peak earlier in high than
in low energies, consistent with a time-of-flight dispersion if the
electrons producing these pulses are accelerated at some distance
from the location where the bremsstrahlung radiation is produced
\citep{asch1996scaling}. However, this does not mean that all
the energetic electrons have to be associated with these sub-second
pulses. The reverse delay in the longer timescale (seconds) HXR
pulses could indicate collisional escape from a coronal trap
\citep{asch1996loop, asch1998deconvolution,
krucker2008HXR}, but could also be a result of a more gradual
acceleration process \citep[e.g.][]{bai1979}.

In this paper we investigate the characteristics of flare emission
across a range of photon energies, and examine the association
between temporal, spatial and spectral characteristics, with
particular interest in the region between thermal and non-thermal
parts of the spectrum.  The paper is organised as follows. We first
review theoretical considerations and present a simple model for the
flare with an isothermal and a power-law X-ray emission component
(Section \ref{theory}). A simple RHESSI flare on 20th February 2002,
with distinct gradual low-energy and impulsive high-energy emissions
is analysed in detail (Section \ref{s:obs}). An overview of the
flare is presented in Section \ref{s:overview}. The semi-calibrated
photon flux is then used to derive the rate of change of photon
fluxes at different energies during a prominent HXR pulse, and two
temporal components are identified (Section \ref{s:semi_ph}). This
is repeated in Section \ref{s:spectral_fit} but using a full
spectral fit. The evolution of model parameters and the
corresponding photon fluxes are used to check self-consistency of
the model, and in Section \ref{s:pivot} we look at the pivot energy
derived from the rate of change of the photon fluxes. In Section
\ref{s:discussion}, we discuss the implications of these results,
and conclusions are drawn in Section \ref{s:conclusions}.

\section{Thermal and Non-thermal X-ray Emission Components of Solar Flares}
\label{theory}

The impulsive phase of most flares is characterised by a
monotonically increasing flux of low-energy emission and a rapidly
varying flux of high-energy emission. RHESSI photon spectra are
usually fitted with an isothermal component at low energies and a
power-law component at high energies.  For the sake of simplicity,
we will ignore details of the radiative processes and assume that
the observed photon flux consists of an isothermal component plus a
power-law component. The photon spectrum therefore is given by
\begin{equation}
 I(\epsilon, t) = I_{th} (\epsilon,  t) + I_{nth} (t)\left({\epsilon}/{{\rm
keV}}\right)^{ - \gamma (t)} \label{equ:model}
\end{equation}
where $\epsilon$ is the photon energy, $I_{th}(\epsilon, t)$ and
$I_{nth}(t)(\epsilon/{\rm keV})^{-\gamma(t)}$ correspond to the
thermal and nonthermal component, respectively. In the presence of a
pivot energy $\epsilon_0$, $I_{nth}(\epsilon_0)=
I_{nth}(\epsilon_0/{\rm keV})^{-\gamma}$ is independent of time $t$
and the variation of the nonthermal component is purely due to
changes in the photon spectral index $\gamma$:  $I_{nth}
(t)\left({\epsilon}/{{\rm keV}}\right)^{ - \gamma (t)} =
I_{nth}(\epsilon_0) (\epsilon/\epsilon_0)^{-\gamma(t)}$.
From the spectral fitting with $I_{th}(\epsilon)$ determined from the full line-plus-continuum spectrum derived from CHIANTI (5.2) as a function of temperature T and emission measure EM, one obtains $I(\epsilon, t_0)$ at a given time indicated by $t_0$. The transition energy $\epsilon_t$
between thermal and non-thermal emissions, where the photon fluxes
produced by the corresponding electron populations are equal, is
determined by $I_{th}(\epsilon_t, t_0)=I_{nth}(t_0)(\epsilon_t/{\rm
keV})^{-\gamma(t_0)}$, where $I_{nth}$ and $\gamma$ are fitting
parameters.

The normalized time rate of change of the photon flux at a given
energy $\epsilon$ is given by
\begin{eqnarray}
R(\epsilon, t) \equiv \frac{{d I(\epsilon, t)}}{{I(\epsilon, t)
\cdot d t}} = \frac{\dot{I}_{th}(\epsilon,
 t)+\dot{I}_{nth}(t)(\epsilon/{\rm keV})^{-\gamma}-
 I_{nth}(t)\dot{\gamma}(t)\ln(\epsilon/{\rm keV}) (\epsilon/{\rm keV})^{-\gamma}}{
 I_{th} (\epsilon, t) + I_{nth} (t)\left({\epsilon}/{{\rm keV}}\right)^{ - \gamma (t)}
 } \label{equ:rate1}  \\
\approx \left\{\begin{array}{ll} R_{th}\equiv
{\dot{I}_{th}(\epsilon, t)}/{{I_{th}(\epsilon, t) }} & \quad \mbox{ {\rm for}
\ \ $\epsilon \leq \epsilon_t$}  \\
R_{nth}\equiv {\dot{I}_{nth}(t)}/{{I_{nth}(t) }} - \dot{\gamma}(t)
\ln
 \left({\epsilon}/{{{\rm
 keV} }}\right) & \quad \mbox{ {\rm for }\ \ $\epsilon > \epsilon_t$}
\label{equ:theory}
\end{array}
\right.
\end{eqnarray}
where the dot above the relevant quantities indicates the derivative
with respect to time and we have used the fact that the thermal and
nonthermal components dominate the low- and high-energy photon
spectra, respectively, to derive the approximate expression. The
rate of change of photon flux of the thermal and power-law component
are indicated by $R_{th}$ and $R_{nth}$, respectively. In the RHESSI
energy range we consider, the thermal continuum spectrum always
dominates at low energies and
the free-free thermal
bremsstrahlung emission is proportional to $EM(t) /$
$\{\epsilon\,T(t)^{1/2}\exp[\epsilon/k_{\rm B}T(t)]\}$, where
$EM(t)= n_{th}^2 V$ and $V$ and $n_{th}$ indicate the source volume and density of the thermal electrons respectively. $k_{\rm B}$ is the Boltzmann constant.
\begin{equation}
R_{th} = {\dot{I}_{th}(\epsilon, t)\over I_{th}(\epsilon, t)} =
{\dot{EM}(t)\over EM(t)} +
 {\epsilon\dot{T}\over k_{\rm B}T^2} -{\dot{T}\over 2T}.
\label{equ:ratet}
\end{equation}
Inclusion of emission lines and free-bound emission will introduce correction terms to this equation.

In principle, one can obtain the right-hand side of Equation
(\ref{equ:rate1}) from the spectral fits. The left-hand side can be
obtained from light curves of different energy bands directly. If
the photon spectral model given by Equation (\ref{equ:model}) is
sufficient, Equation (\ref{equ:rate1}) should be satisfied. If the
change in the temperature of the thermal component is small, then
the rate of change of the photon flux is independent of the photon
energy $\epsilon$ in the thermally-dominated energy range (Eq.
\ref{equ:ratet}). One can therefore obtain $R_{th}$ by fitting the
rate of change of photon flux  in the thermal regime with a function
independent of $\epsilon$. There will be an energy
$\epsilon^\prime_t$, where the time derivatives of thermal and
non-thermal components are equal, i.e.
$R_{th}=R_{nth}(\epsilon^\prime_t)$ and if the simple equivalence
between non-thermal/thermal emission and impulsive/gradual emission
holds  then $\epsilon^\prime_t$ should be comparable to $\epsilon_t$
obtained from spectral fits. Equation (2) also provides a means to
investigate the presence of a constant `pivot' energy $\epsilon_0$ for the
power-law component, which will occur where $R_{nth}$ goes to zero -
the invariant point in the photon spectrum of the power-law
component.

\section{Observations}
\label{s:obs}

RHESSI observed a flare on 20th February 2002 in the NOAA active
region 9825, located near the northwest limb of the Sun at N16W80
(919$\arcsec$W, 285$\arcsec$N) a few days after its successful
launch on 5 February 2002. It is the focus of several earlier
studies dealing with the characteristics during the HXR peak
\citep{sui2002feb20,asch2002chromos}. The imaging and
spectroscopic software has been improved significantly since then
and the instrumental response is better understood and incorporated
in the RHESSI software packages. We choose this event because
of the very simple shape of its light curves, and present an
investigation here emphasizing the relationship between the low-energy
and high-energy emission components.

\subsection{Light curves, Images, and Spectra}
\label{s:overview}

\begin{figure}
\begin{center}
\includegraphics[width=0.8\textwidth]{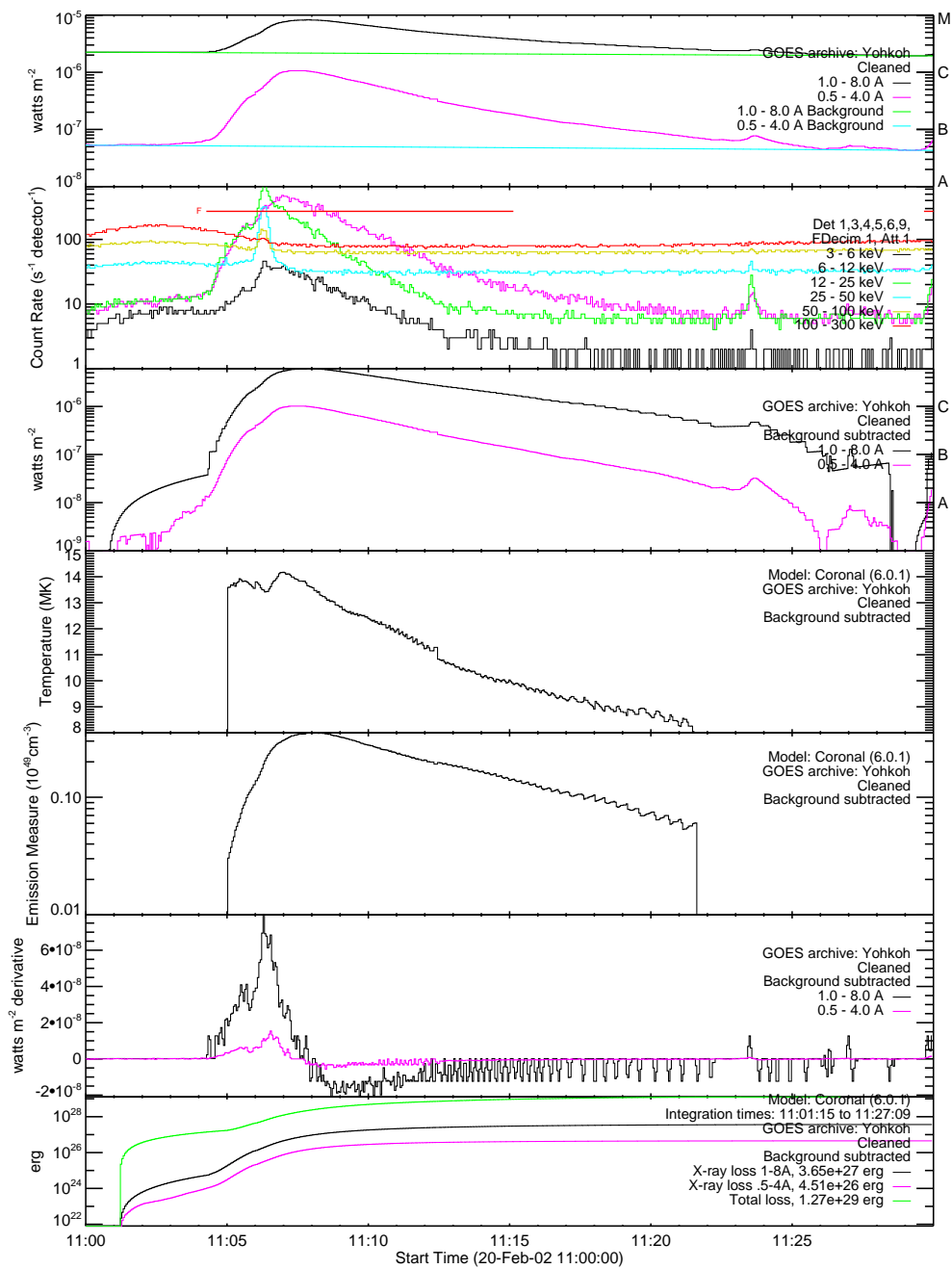}
\end{center}
\caption{Summary of RHESSI and GOES observations. The high-lightened range shows the time interval (11:05:50 ---11:06:50) used for images and spectra in Figures ~\ref{fig:image2} and~\ref{fig:image3}.  The top panel
shows the soft X-ray (SXR) lightcurves and chosen backgrounds (as linear interpolation of fluxes of two intervals before and after the flare) for both energy bands observed by GOES.
 The second panel shows the RHESSI count rate in several energy bands. The third panel shows GOES lightcurves with the background subtracted. The fourth and fifth panels show, respectively, the temperature and emission measure derived from the GOES fluxes, in agreement with results obtained by
 Sui et al. (2002). The sixth panel shows the time-derivatives of the
 GOES lightcurves, which is qualitatively correlated with the HXR count
 rates in the top panel. The bottom panel shows the energies radiated by the hot plasma observed with GOES obtained using CHIANTI 6.0.1 with coronal abundances. }
 \label{fig:goes}
\end{figure}

This GOES 1-8 \AA~  C7.5 flare exhibits
a prominent HXR pulse lasting for about 50 seconds with count rates
above 12 keV peaking near 11:06:20 UT. The first three panels of Figure \ref{fig:goes} show a
summary of GOES and RHESSI observations of this flare. The 3-6~keV
light curve shows some impulsive behavior, while the 6-12~keV counts
are relatively smooth. The RHESSI attenuator state during this flare
was A1.
Counts in the 3-6 keV channel when the attenuators are in place are almost all from higher energy photons above 11 keV, because of the effect of K-escape \citep{smith2002rhessi}. We therefore only analyze counts above 6 keV in this work.
The attenuator also reduces the count rate at low energies significantly leading to a livetime better than 93\% for all the detectors. Pulse pileup can then be ignored in the spectral study \citep{smith2002rhessi}. The rise of the RHESSI count rates below 25 keV becomes evident after 11:04 UT marking the onset of the flare.
The slowly rising count rates before 11:04 UT are likely caused by particle events as is evident from the gradual varying count rates in higher energy channels where the statistical errors are significant. The background profile can be subtracted with sufficient accuracy by modeling this gradual varying component in different energy channels separately.

The fourth and fifth panels of Figure \ref{fig:goes} show the GOES temperature and emission measure, respectively.
The background fluxes in the two energy channels are chosen as a linear interpolation between average fluxes during two intervals before and after the flare and are shown in the first panel. The temperature and emission measure can not be obtained before 11:05 UT, presumably due to the relatively low background subtracted fluxes shown in the third panel. The values obtained for the impulsive phase are rather insensitive to the background selection. These results are in agreement with those obtained by \citet{sui2002feb20} for the prominent HXR pulse. We note that the temperature does not change significantly throughout the rise phase of the 6-12 keV count rate from 11:05 to 11:07 UT. The emission measure appears to grow exponentially at the beginning with a growth time of $\sim 40$ seconds, and the growth rate decreases significantly after the HXR pulse. The sixth panel shows the time derivatives of the GOES fluxes. Although these derivatives peak near the peak of the HXR pulse, and a secondary peak before the major peak appears to be correlated with RHESSI light curves below 25 keV in rough agreement with the Neupert effect, a broader correlation is not very obvious. Since the first peak does not contain high energy ($>$ 25 keV) emissions which normally have a longer decay time than lower energies, we assume that the two peaks are independent and the first peak does not significantly affect the spectral properties determined for the major one.
The bottom panel shows the radiative energy produced by the isothermal source, obtained by fitting the GOES fluxes using CHIANTI
6.0.1 and assuming coronal abundances \citep{dere2009}. The total radiated energy from the hot plasma is about $10^{29}$ ergs for this flare.

\begin{figure}[]
\begin{center}
\includegraphics[width=0.47\textwidth]{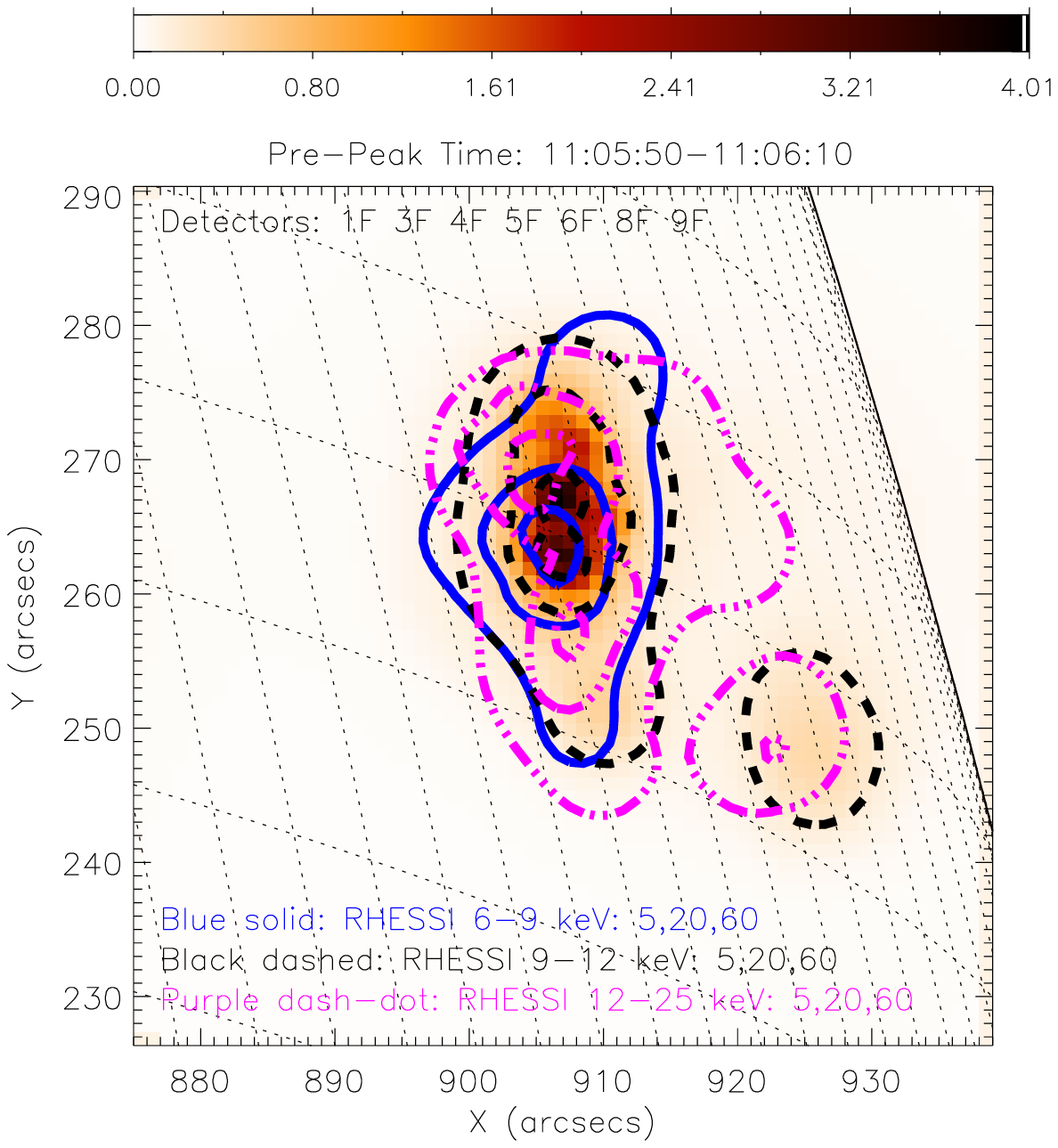}
\includegraphics[width=0.47\textwidth]{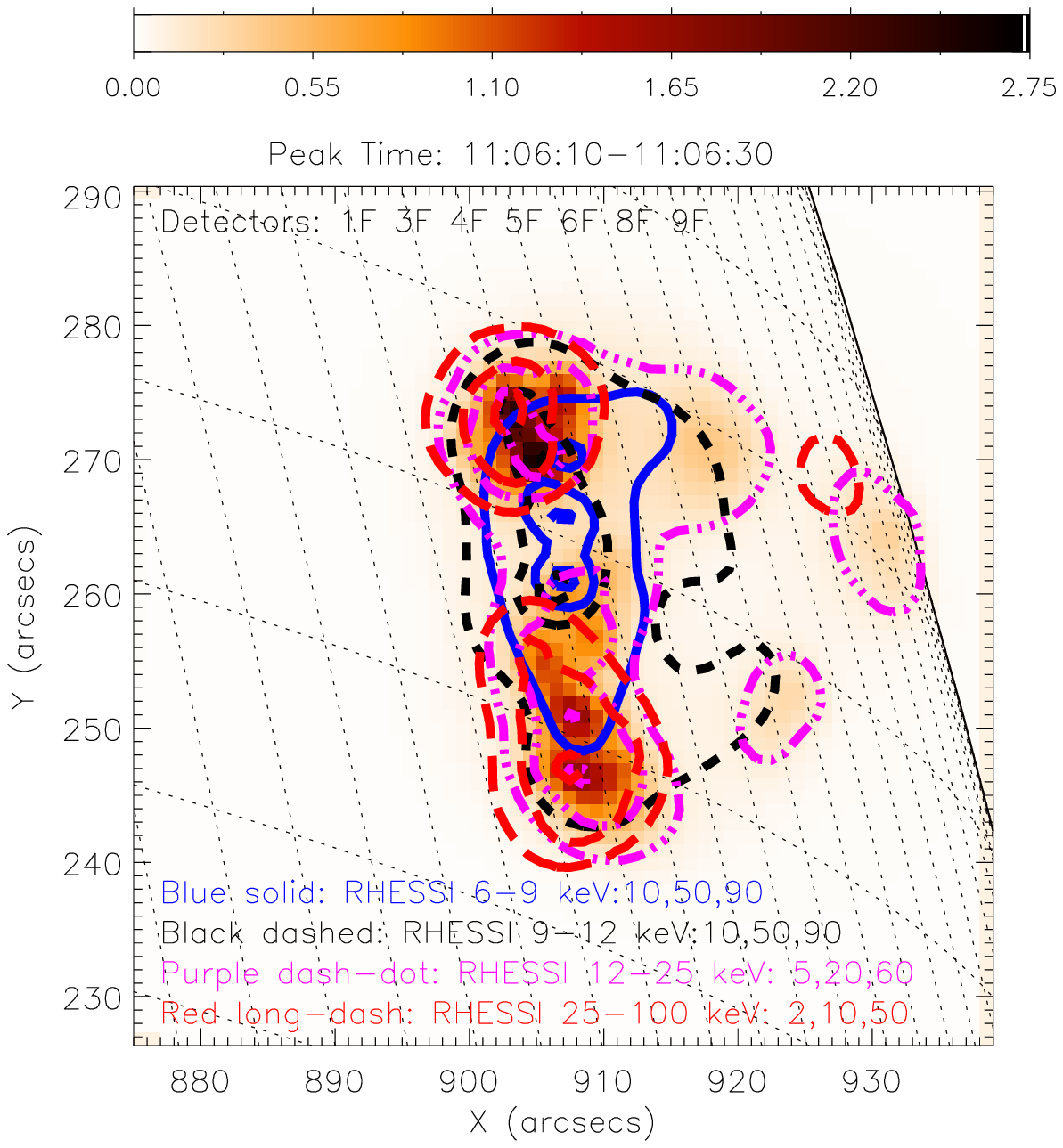}
\end{center}
\caption{Images of the flare for a 20 second interval before (Left)
 and a 20 second interval during (right) the HXR peak. The map color is in the unit of photons ${\rm{cm}}^{-2}$ ${\rm{s}}^{-1}$ ${\rm{arcsec}}^{-2}$.
 {\it Left:} The image and dashed contours (5, 20, and 60 \% of
 the peak brightness) are for 9-12 keV energy band. The solid (5, 20, and 60 \%) and dash-dot (5, 20, and 60
 \%) contours are for the 6-9 keV and 12-25 keV energy band,
 respectively.
 {\it Right:} The image and dot-dash contours (5, 20, and 60 \%) are
 for the 12-25 keV energy band.
 The solid (10, 50, and 90 \%), dashed (10, 50, and 90\%)
 and long-dash (2, 10, 50\%) contours are for the 6-9 keV, 9-12 keV
 and 25-100 keV energy band, respectively.} \label{fig:image2}
\end{figure}

\begin{figure}[ht]
\begin{center}
\includegraphics[width=0.32\textwidth]{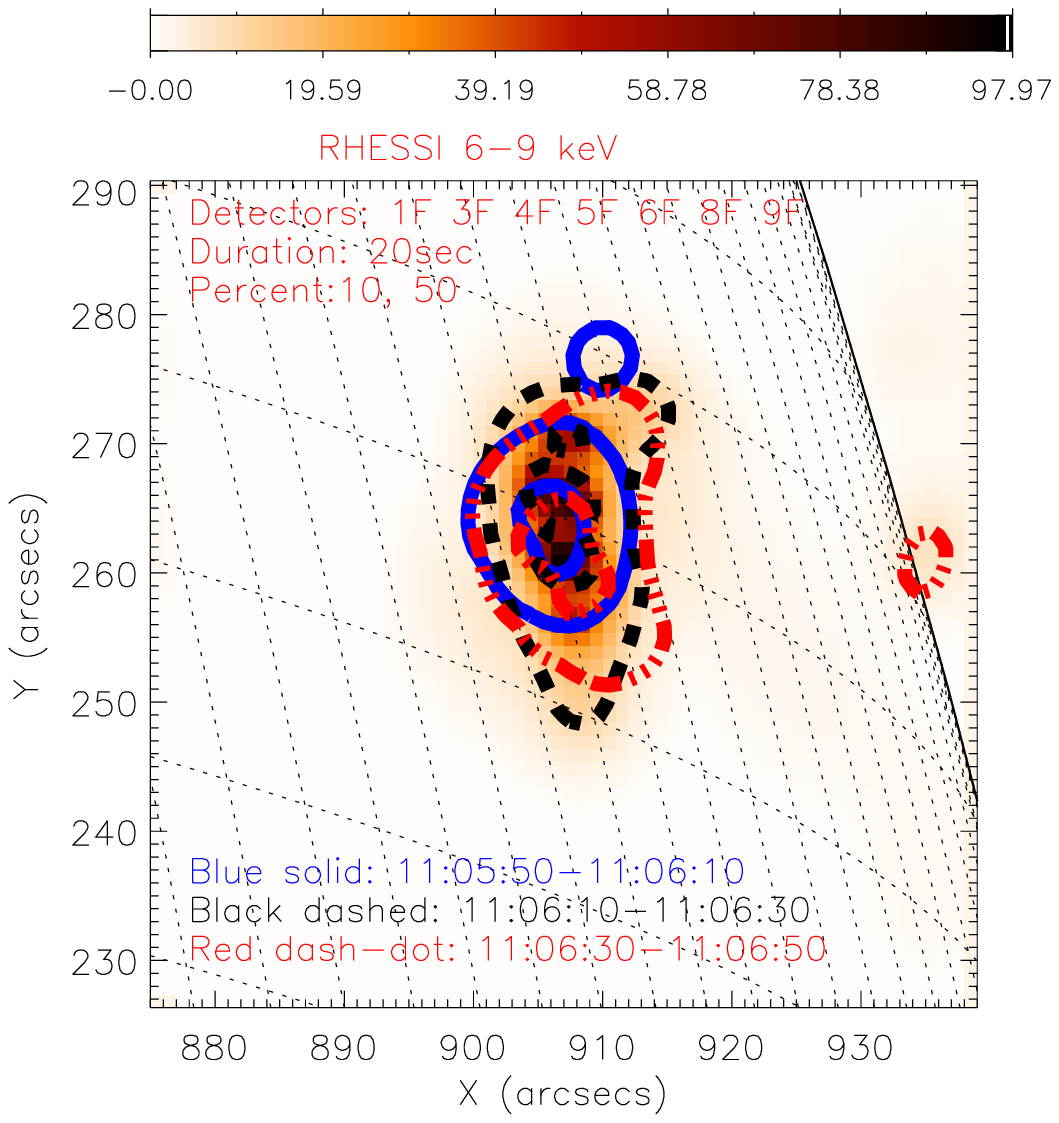}
\includegraphics[width=0.32\textwidth]{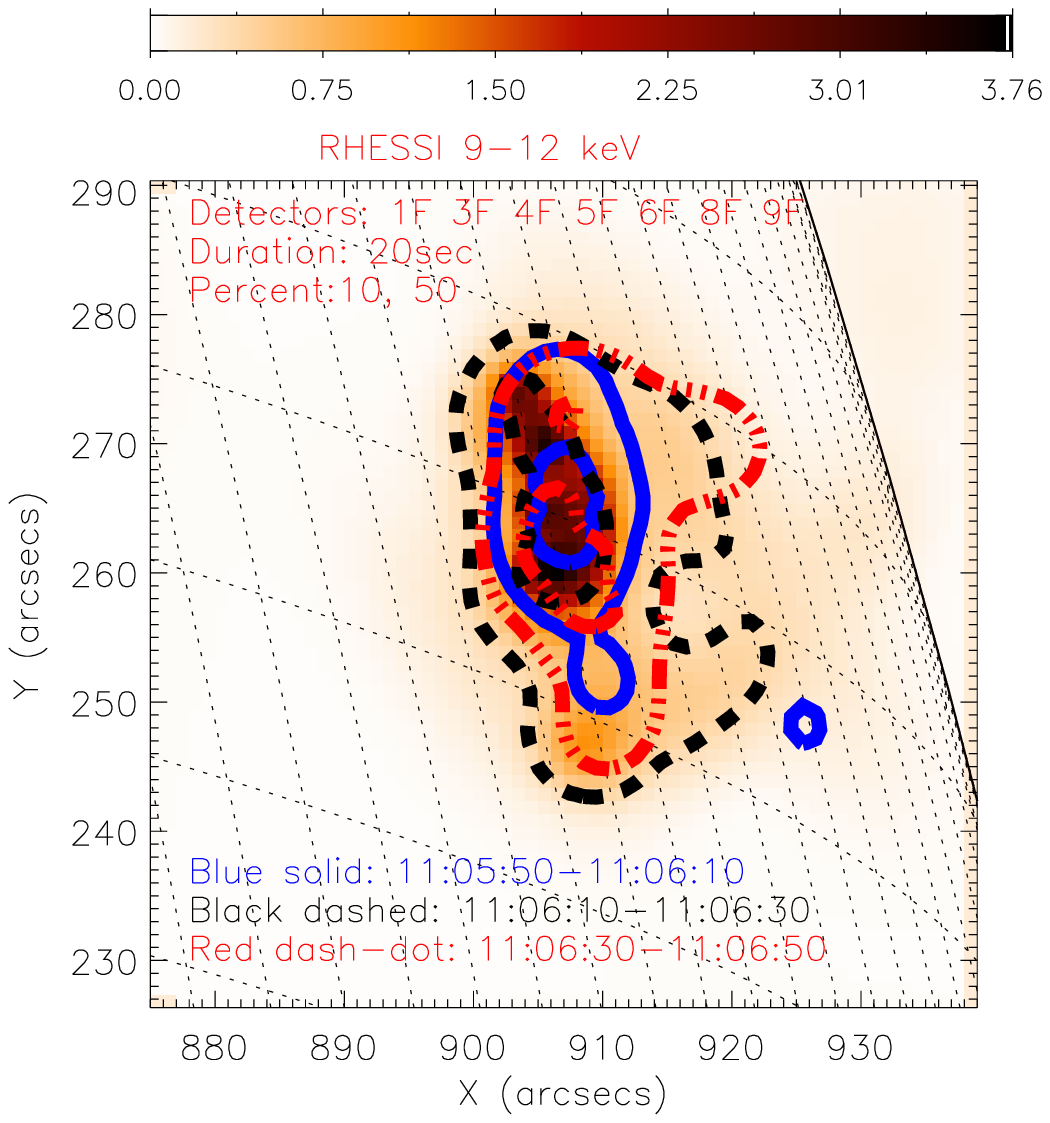}
\includegraphics[width=0.32\textwidth]{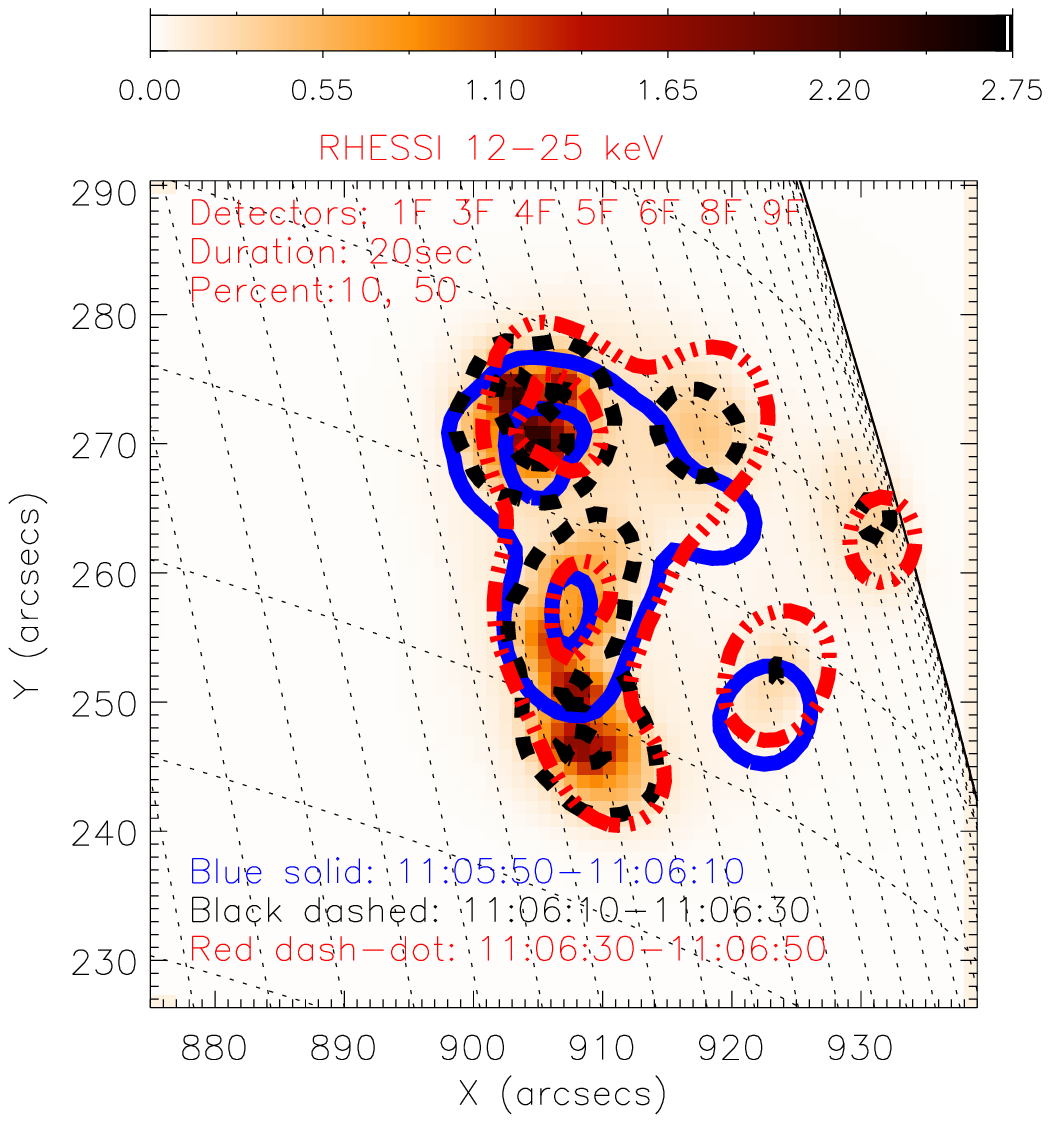}
\end{center}
\caption{Evolution of the X-ray images at 6-9~keV (left), 9-12~keV
(middle), and 12-25~keV (right). The map color is in the unit of photons ${\rm{cm}}^{-2}$ ${\rm{s}}^{-1}$ ${\rm{arcsec}}^{-2}$.
 The image and dashed contours (10, 50\%) are for the interval of the HXR peak.
 The solid contours and dash-dot contours are for the interval
 before and after the HXR peak, respectively.
} \label{fig:image3}
\end{figure}

The right panel in Figure \ref{fig:image2} shows the source structure
at several energies during the prominent HXR peak from 11:06:10 to
11:06:30 UT obtained using the Pixon algorithm \citep{Pina1993}.  The weak HXR coronal source near the solar limb has been interpreted as the site of particle acceleration by \citet{sui2002feb20}.
\citet{asch2002chromos}, on the other hand, inferred a much
smaller loop. Our results suggest that both large and small loops
are present at energies up to 25~keV. There are also clear
footpoints at $>$ 25~keV which appear to be at the end of the lower
energy loops. This imaging does not support the suggestion by Sui et
al. (2002) that there is an independent low-energy thermal source
between the two HXR footpoints, rather it indicates that the small
loop connects to the HXR footpoints. The structure just before the
flare peak is shown in the left panel of Figure \ref{fig:image2}.

The time evolution of X-ray sources at 6-9~keV, 9-12~keV, and
12-25~keV are shown in Figure \ref{fig:image3}. The 6-9 keV source
structure (in the left panel) is relatively simple, loop-like and
compact before the HXR pulse. It becomes more extended in the
following two time intervals and develops three sub-sources (though
there is a possibility that this is over-resolution by the Pixon
algorithm). The 9-12~keV emission is primarily loop-like but more
extended than its 6-9~keV counterparts. Features associated with the
footpoints start to emerge.
The distinction between footpoint sources and loop source(s) becomes
very ambiguous at 12-25 keV (in the right panel of Figure
\ref{fig:image3}). The brightest locations are associated with the
footpoints at the HXR peak, but the 12-25 keV structure is elongated
(as is the 9-12~keV source in the middle panel) and may
include a loop component.  The X-ray images therefore reveal a
complicated pattern of low and high energy sources, with no clear
distinction between footpoints and loop at energies from $\sim$10 to
25~keV.

\begin{figure}[ht]
\includegraphics[width=0.32\textwidth]{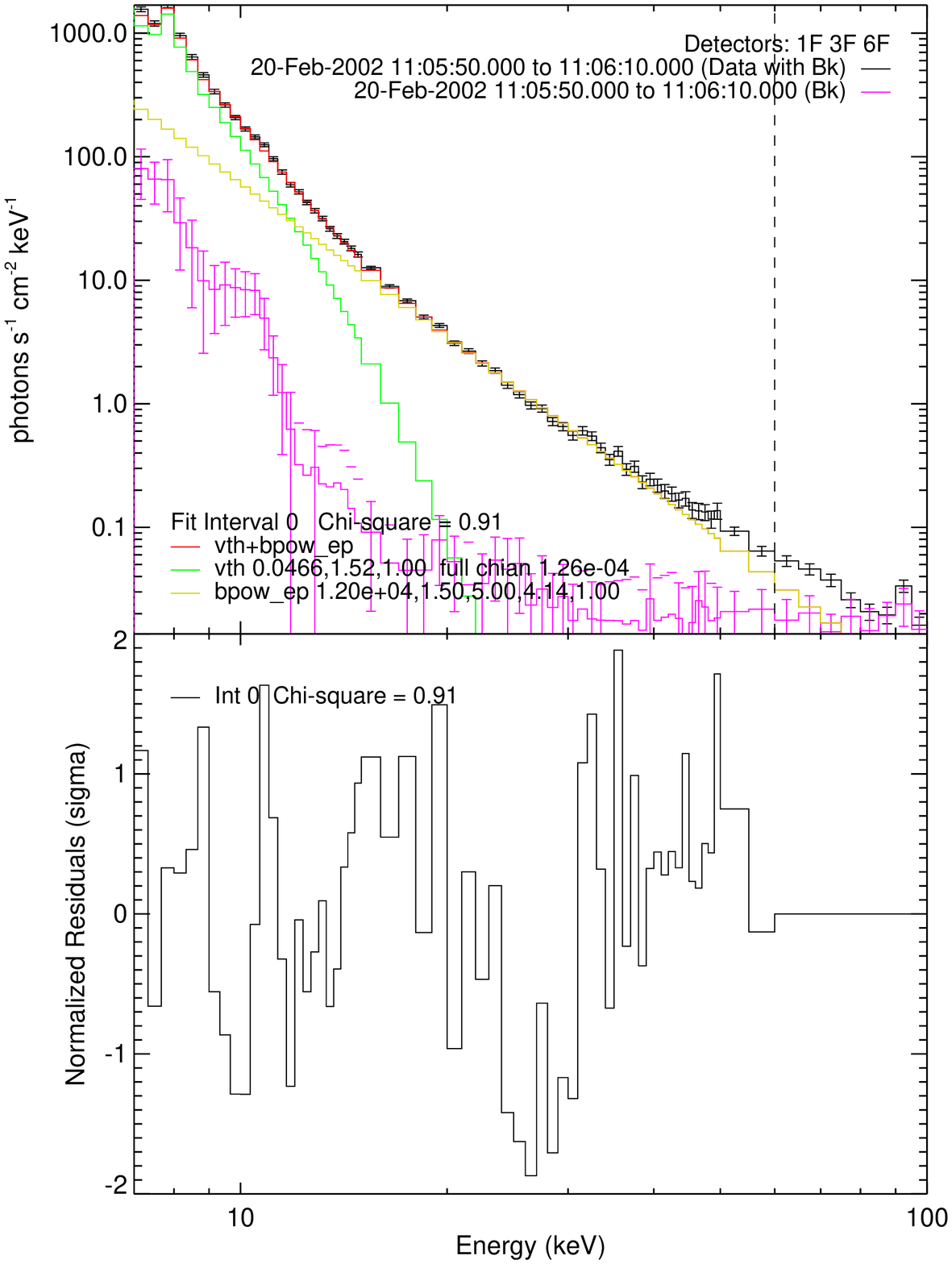}
\includegraphics[width=0.32\textwidth]{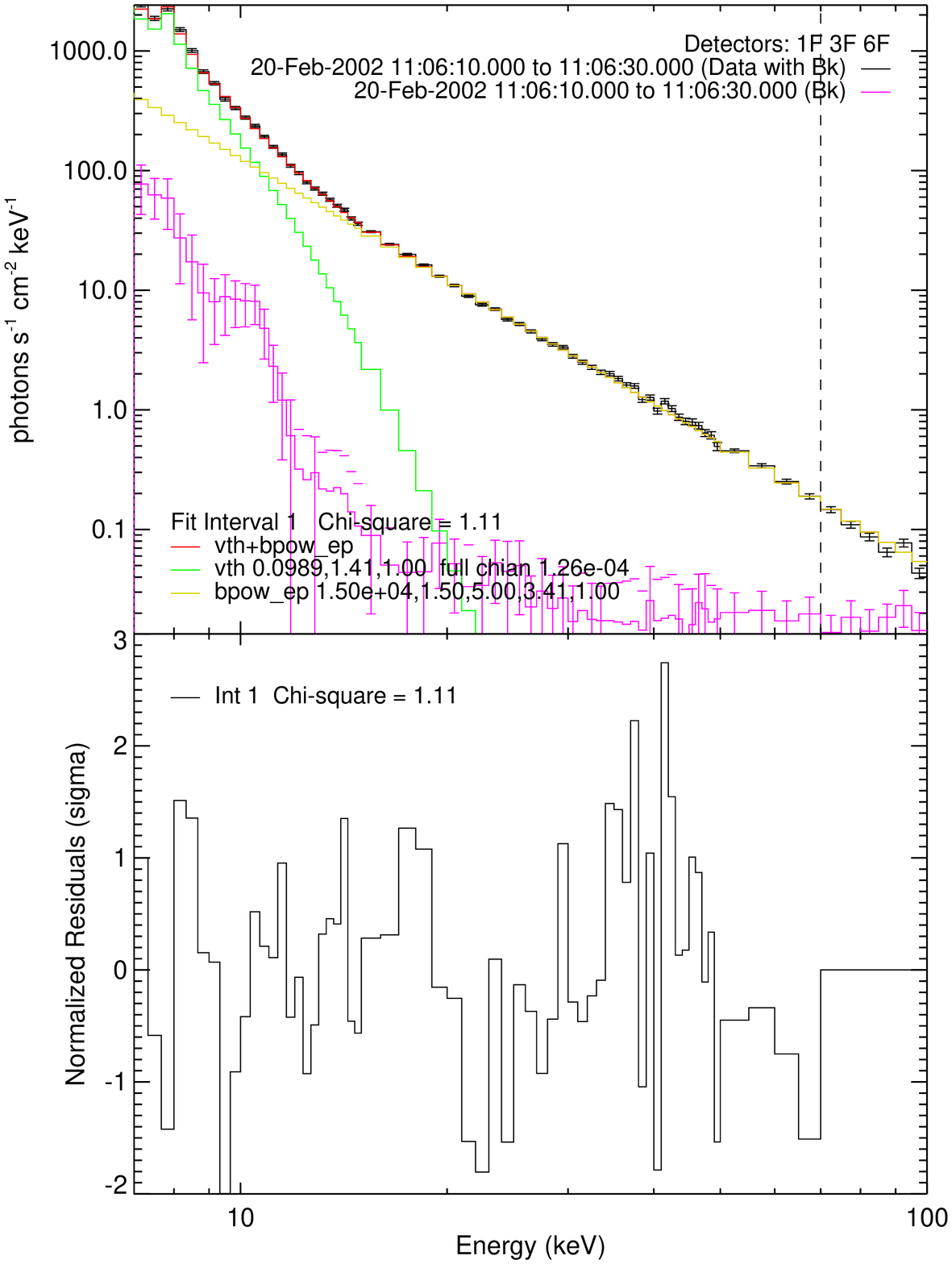}
\includegraphics[width=0.32\textwidth]{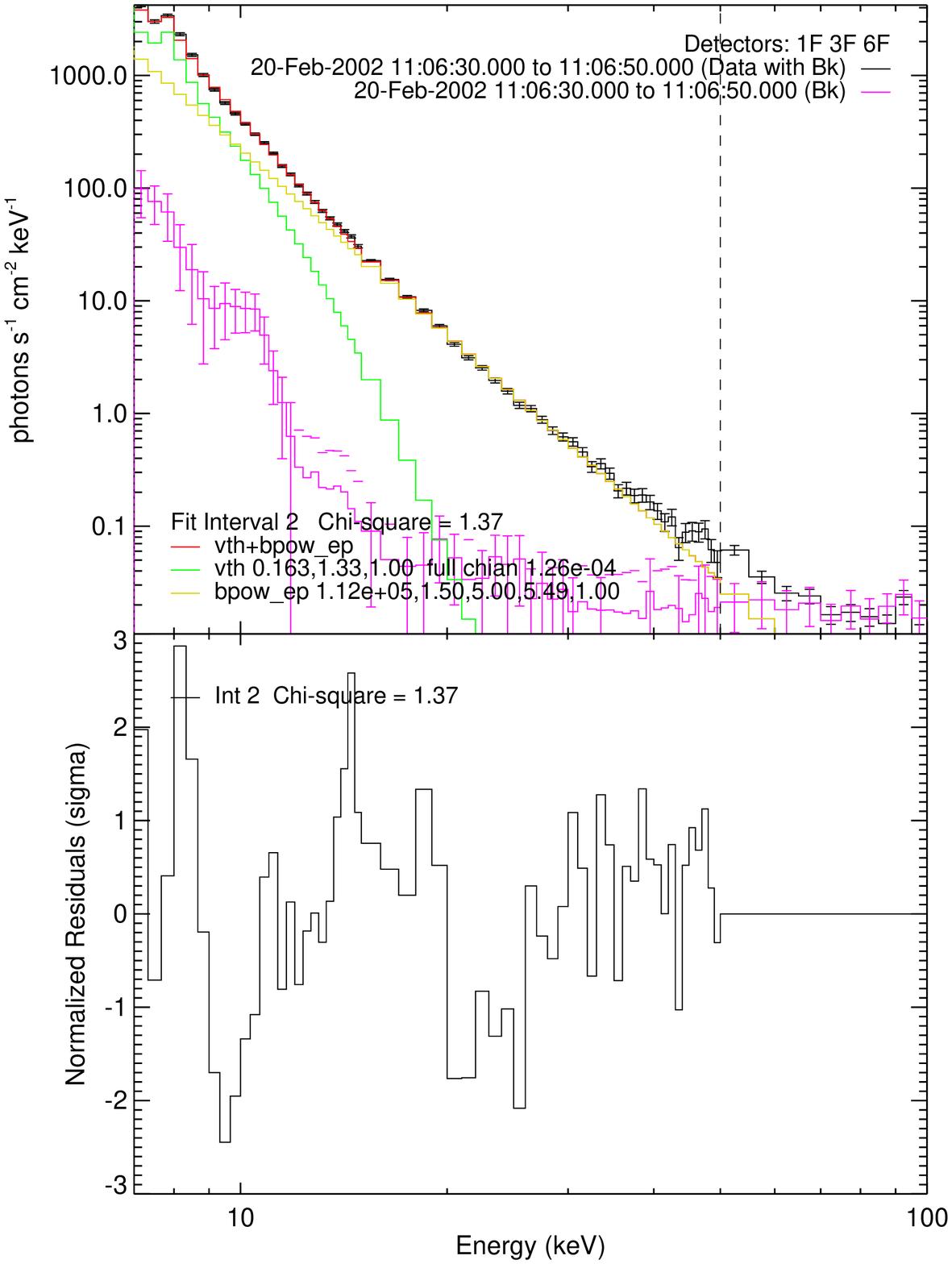}
\caption{Photon spectra of the three 20-second intervals in Figure
 \ref{fig:image3} fitted with an isothermal plus a power-law model.
 Time starts from 11:05:50 and increases from left to right.
 Model parameters are indicated in the figure.} \label{fig:spec}
\end{figure}

Figure \ref{fig:spec} shows the photon spectra fitted with an
isothermal plus a power-law component for the three 20-second
intervals of Figure \ref{fig:image3}.  For this preliminary study, the RHESSI background is chosen as a linear interpolation, made between the average counts during the intervals 11:01:58 - 11:02:38 and 11:20:38 - 11:21:18. More detailed modeling of the background is carried out in Section~\ref{s:spectral_fit}, where the spectra for 4-second intervals are analyzed. (Although we choose a broken
power-law model in the fitting, the break energy is fixed at 5 keV
which is below the energy range of the data.) The soft-hard-soft
spectral evolution is evident with the photon spectral index varying
from 4.1 to 3.4 and to 5.5. The emission measure and temperature are
$EM = 4.7\times10^{47}$ cm$^{-3}$, $kT=$1.5 keV;  $EM =
9.9\times10^{47}$cm$^{-3}$, $kT =$1.4 keV, and $EM =
1.6\times10^{48}$ cm$^{-3}$, $kT =$1.3 keV, respectively. The emission measure is slightly lower and the temperature is slightly higher than those from the GOES spectral fit, which may be attributed to the different energy ranges covered by these two instruments.
Comparison of these spectra shows that the 12-25 keV emission is more and more
dominated by the power-law component as the flare evolves.
The spectral fit of the last interval has also the highest values of
reduced $\chi^2$ (1.37) and residuals, leading to a probability of 8\% to get a larger $\chi^2$ assuming a correct model. Indeed, for the last interval a thermal plus a broken power-law model gives much improved spectral fit. We also study the spectral evolution after the impulsive phase. The simple thermal plus a single power-law model can be ruled out by the relatively softer high-energy spectra, which imply dominance of very low-energy emission by the power-law component. A thermal plus a broken power law or multi-thermal model can give acceptable fit.
However, one should note that the high-energy photon fluxes change dramatically during the two intervals before and after the HXR peak and the large systematic residuals around the iron-line complex at 6.7 keV indicate that this feature has not been modeled properly. Currently, the only way to improve the modeling of emission lines is to fit spectra from individual detectors and take into account small gain changes and pulse pile up, which is beyond the scope of the current investigation.

\subsection{Semi-calibrated light curves and their rate of change}
\label{s:semi_ph}

\begin{figure}
\centerline{\includegraphics[width=0.8\textwidth]{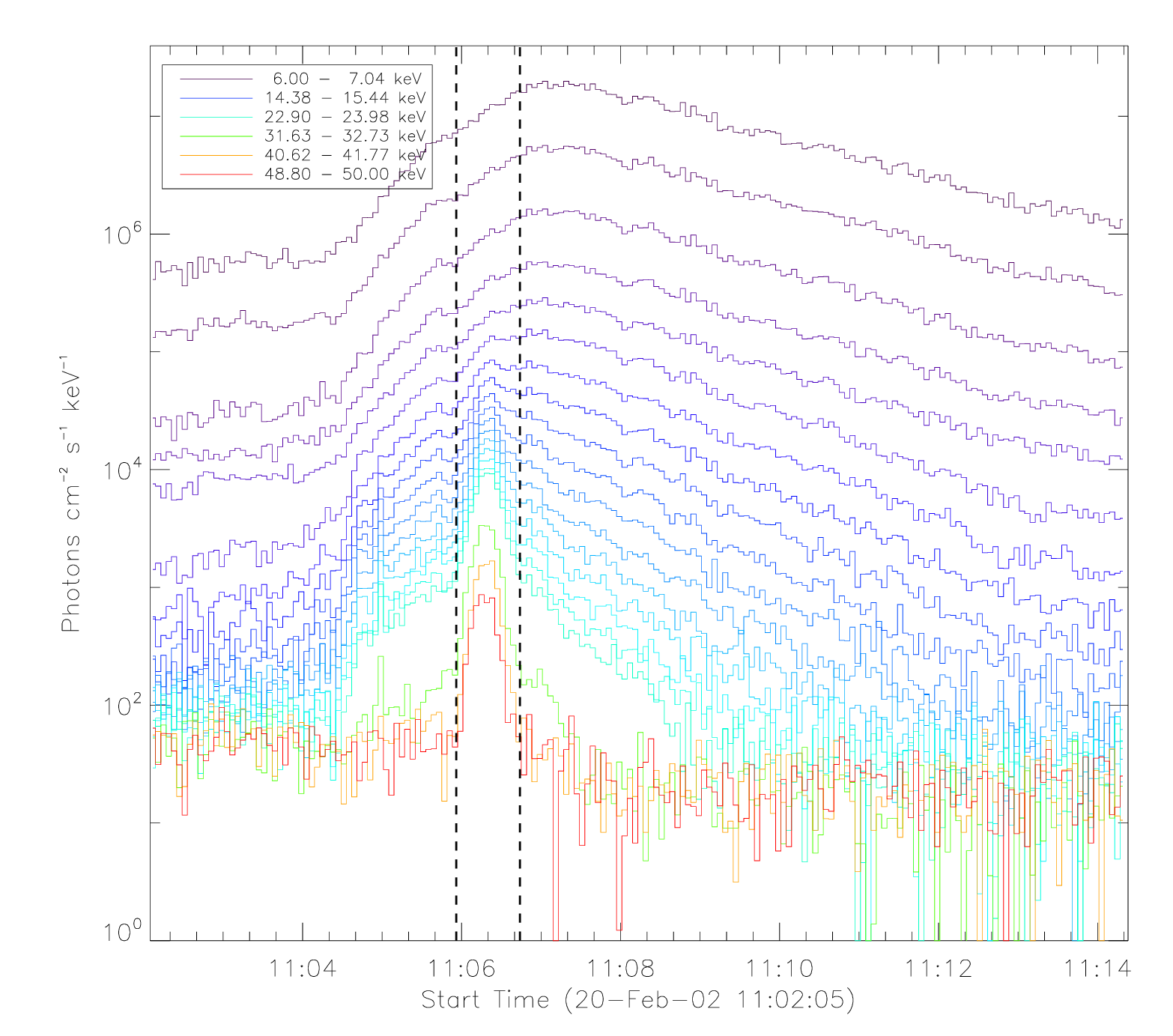}}
\caption{Semi-calibrated photon flux $f(\epsilon,t)$ of the flare averaged over 4 s intervals. The curves with different
colors represent photon counts in 40 different energy bands from 6 to 50 keV. The vertical dash lines represent the start (11:05:56) and end time (11:06:44) of the HXR pulse. Notice that not all of the 40 energy bands are shown.} \label{fig:2022003_photon0}
\end{figure}

\begin{figure}
\centerline{\includegraphics[width=0.8\textwidth]{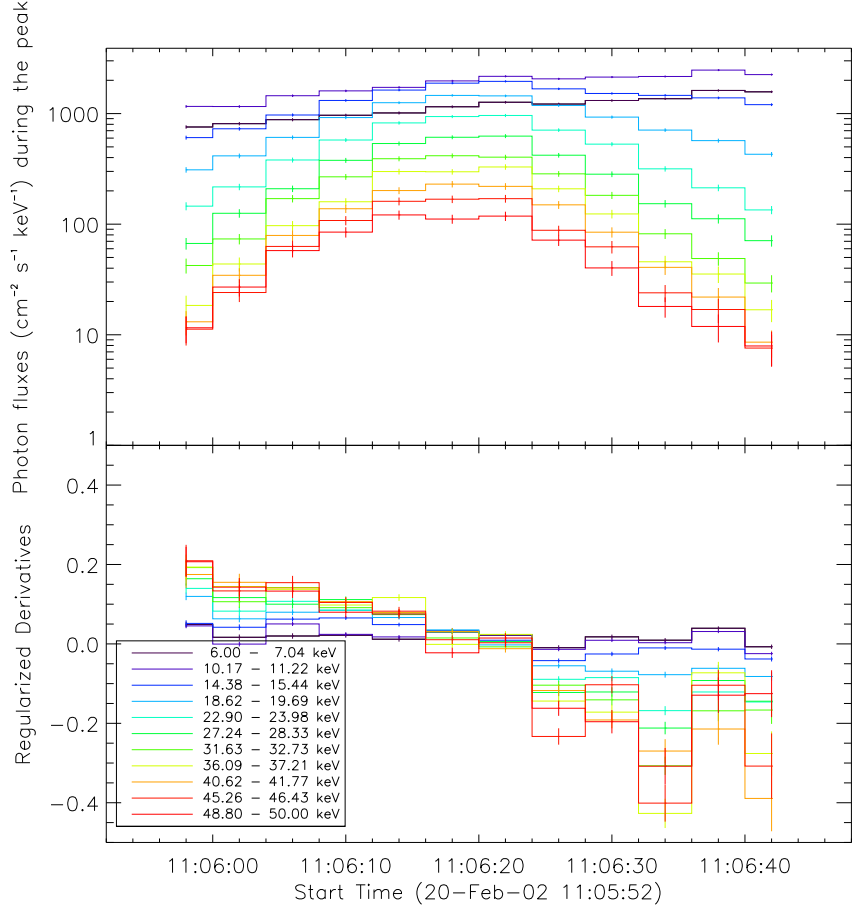}}
\caption{\emph{Top}: Semi-calibrated photon flux $f(\epsilon,t)$ and its $\pm1 \sigma$ statistical errors during the HXR pulse (11:05:56---11:06:44). \emph{Bottom}: The rate of change of the photon flux $R(\epsilon,t)= df(\epsilon,t)/dt/f(\epsilon,t)$ ($\rm{s^{-1}}$) and its corresponding $\pm1\sigma$ errors calculated with the regularized method. Notice that we have shown only 11 out of the 40 energy bands.} \label{fig:2022003_PHdrv}
\end{figure}

\begin{figure}
\centerline{\includegraphics[width=0.8\textwidth]{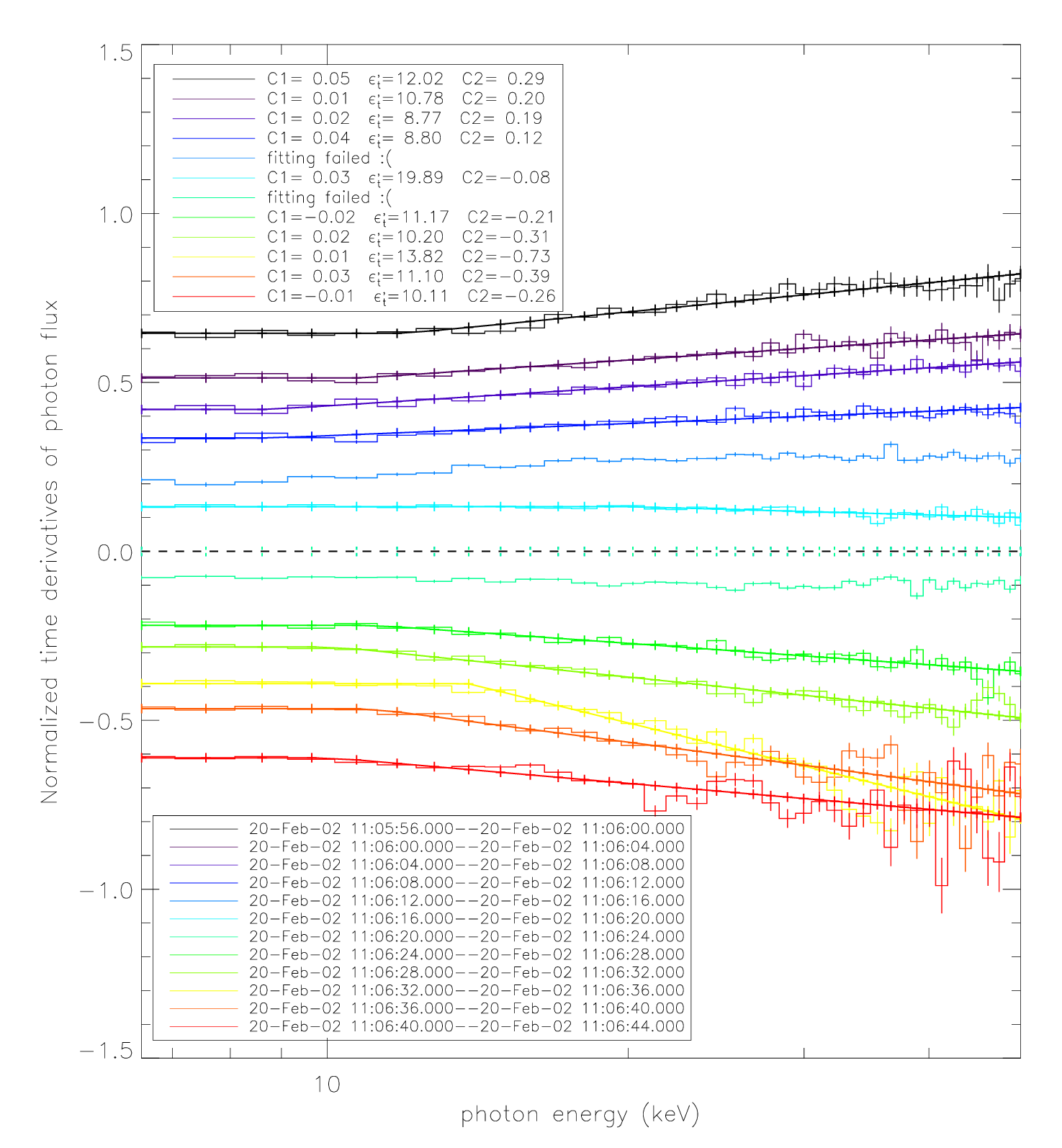}}
\caption{Rate of change of semi-calibrated photon flux $R(\epsilon,t)= df(\epsilon,t)/dt/f(\epsilon,t)$ ($\rm{s^{-1}}$) versus the
photon energy for 12 4-second time bins. The error bars give $\pm 1\sigma$ uncertainties. The data are fitted
with the model described by Equation (\ref{equ:DRV_E_fit}) and the best
fit models are shown as solid lines. For illustrative purpose, the lines are shifted vertically by the values of [0.6, 0.5, 0.4, 0.3, 0.2, 0.1, -0.1, -0.2,
-0.3, -0.4, -0.5, -0.6] in time sequence. The fit does not converge for
the fifth and seventh time intervals. The model parameters C1,
$\epsilon_t$ and C2 of the fit are indicated in the legend.  }
\label{fig:2022003semi_drv_e}
\end{figure}

We here study the photon flux change rate $R(\epsilon,t)$ given by Equation
(\ref{equ:theory}) to quantify the rate of change of photon flux ---
how impulsive or gradual the event is --- as a  function of both
energy and time. We do this first in a model-independent way using
the semi-calibrated photon flux. Semi-calibrated photon fluxes can
be obtained from the observed count rates by using the diagonal
elements of the spectral response matrix. Although the photon flux
obtained this way does not take into account the full spectral
response matrix of RHESSI, it can be readily obtained and gives an
approximate description of the photon flux from the source, and one
which does not depend on assuming a particular form of the photon
spectrum (note, we investigate full spectral fitting in
Section~\ref{s:spectral_fit}). Following the arguments of Section
\ref{theory}, with the photon spectrogram obtained this way, one can
test whether the two components identified from spectral fits are compatible with the two temporal components.

Figure ~\ref{fig:2022003_photon0} shows some of the detailed semi-calibrated photon fluxes $f(\epsilon,t)$ in forty energy
bands between 6 and 50 keV. Each energy bin is set to be no smaller than 1 keV which is the energy resolution of RHESSI \citep{smith2002rhessi}. The
vertical axis indicates the photon flux averaged over a 4 second
interval in the corresponding energy band.
The background fluxes of high energy bands decrease gradually with
time. We model these background fluxes for different energy bands
with a first order polynomial fitting of background values obtained
before and after the HXR pulse.

We use the regularized method developed by
\citet{kontar2005regularized} to obtain the time derivatives of
these light curves. This method gives smoother derivatives while
avoiding large errors, typical of finite differences of discrete
numerical data.
We assume statistical error for the photon flux $\sigma f(\epsilon,t) = M^{-1}(\epsilon)\ {C(\epsilon,t)}^{1/2}$ where $M$ is the diagonal components of the instrument response matrix and $C$ is the count rate. The top panel in Figure ~\ref{fig:2022003_PHdrv} gives the photon flux $f(\epsilon,t)$ and its $1\sigma$ statistical error $\sigma f(\epsilon,t)$ in different energy bands during the HXR pulse.
The bottom panel shows the rate of change of photon flux $R(\epsilon,t)= df(\epsilon,t)/dt/f(\epsilon,t)$ and its $1\sigma$ error $\sigma R(\epsilon,t)$, which is the standard deviation of $R(\epsilon,t)$ modeled with a Gaussian by sampling 5000 points in $df(\epsilon,t)/dt$ and $f(\epsilon,t)$ within their respective $1\sigma$ range of a Gaussian distribution. Here only the $1\sigma$ range of a Gaussian distribution is sampled for the following reason. The relatively low flux $f$ and its relatively large $1\sigma$ error mean that a sampling over the full distribution will have points with $f(\epsilon,t)$ close to zero. This will lead to very high values of $R(\epsilon,t)$, whose distribution is poorly fitted with a Gaussian. The $1\sigma$ error obtained this way should be considered as a lower limit.
The photon fluxes at lower energies have a more gradual temporal
evolution and lower rate of change, in contrast to the high-energy
band photon fluxes, which have rapid rise and decay phases and
highly variable rate of change. The absolute value of the rate of
change often increases with the photon energy. However, the $1\sigma$ error of the rate of change of photon flux also increases with energy, and the variation in the rates of change at high energies may not be significantly different from those at low energies. At the time bin for
the peak from 11:06:20 to 11:06:24, however, the rates for all
energies are around zero.

Figure~\ref{fig:2022003semi_drv_e} shows the energy dependence of
the rate of change of photon flux at different 4 second time
intervals of the HXR pulse. It is clear that the higher energy
fluxes have higher values of the derivative in the rise phase
(before 11:06:20) and lower values of the derivative in the decay
phase (after 11:06:24). This confirms that higher energy fluxes are
more variable than those at low energies. There appears to be two
temporal components. At low energies, the rate of change is nearly
independent of the photon energy. At high energies, the energy
dependence of the rate of change appears to increase linearly with
the logarithm of the photon energy. To quantify these results, we
adopt the following model for the rate of change R
\begin{eqnarray}
R= \left\{\begin{array}{lcr}
 C1 & {\rm for } &   \epsilon \leq \epsilon^\prime_t \\
 C1 + C2 \log_{10} ({\epsilon} /{\epsilon^\prime_t}) & {\rm for } & \epsilon > \epsilon^\prime_t.
 \label{equ:DRV_E_fit}
\end{array}
\right.
\end{eqnarray}
There are therefore three model parameters: C1, C2, and
$\epsilon^\prime_t$. The solid lines in Figure
\ref{fig:2022003semi_drv_e} indicate the best-fit model (using the
curvefit function in IDL). Note that the 4th and 7th time intervals
are not fitted with the model due to significant uncertainty in
$\epsilon^\prime_t$ because of low values of the rate of change of
photon flux at all energies near the HXR peak.

A comparison of equations~(\ref{equ:theory}) and
(\ref{equ:DRV_E_fit}) shows that
\begin{eqnarray}
 C1 &= &R_{th} = {\dot{EM}(t)}/EM(t)\,,\label{equ:C1}\\
 C2 &=& - \ln 10 \cdot \dot {\gamma}\,,    \label{equ:C2}
\end{eqnarray}
where we have assumed that the temperature of the thermal component
does not change during the HXR pulse (demonstrated using spectral
fitting in the next Section). $\epsilon^\prime_t$ is the transition
energy, where the rate of change of the high and low energy
components are equal: $R_{th}=R_{nth}(\epsilon^\prime_t)$. If the
isothermal and power-law model of equation (\ref{equ:model}) indeed
gives sufficient description of the observations,
$\epsilon^\prime_t$ should be comparable to the transition energy
identified from the spectral fit $\epsilon_t$.

\subsection{Spectral fit and the rate of change of photon flux and model
  parameters}
\label{s:spectral_fit}

\begin{figure}
\centerline{\includegraphics[width=\textwidth]{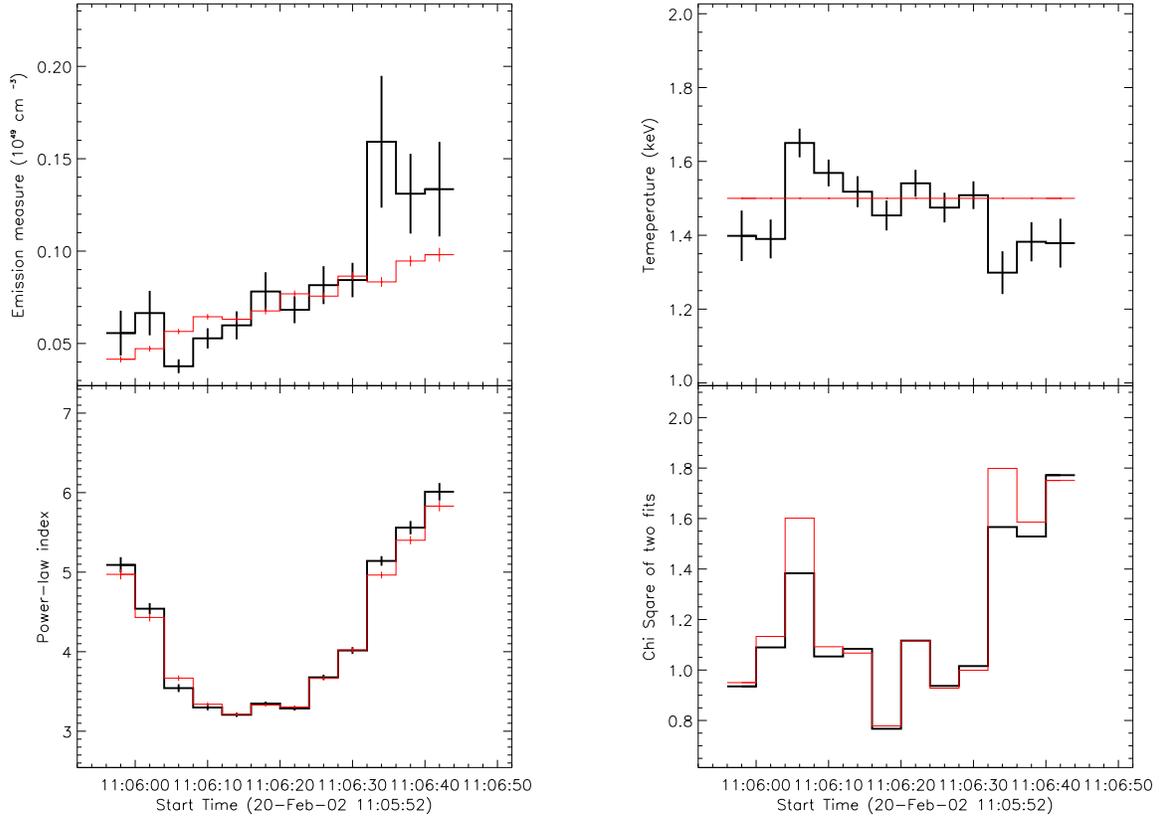}}
\caption{Parameters of an isothermal plus a single power-law
spectral fits with temperature adjustable (thick black lines) and
temperature fixed at 1.5 keV (thin red lines). The top-left, top-right,
bottom-left,
 bottom-right panels show the emission measures, temperatures, the
 power-law spectral indexes, and the ${\chi}^{2}$ of both fits,
 respectively.
} \label{fig:2022003para_spec}
\end{figure}

\begin{figure}
\centerline{\includegraphics[width=0.8\textwidth]{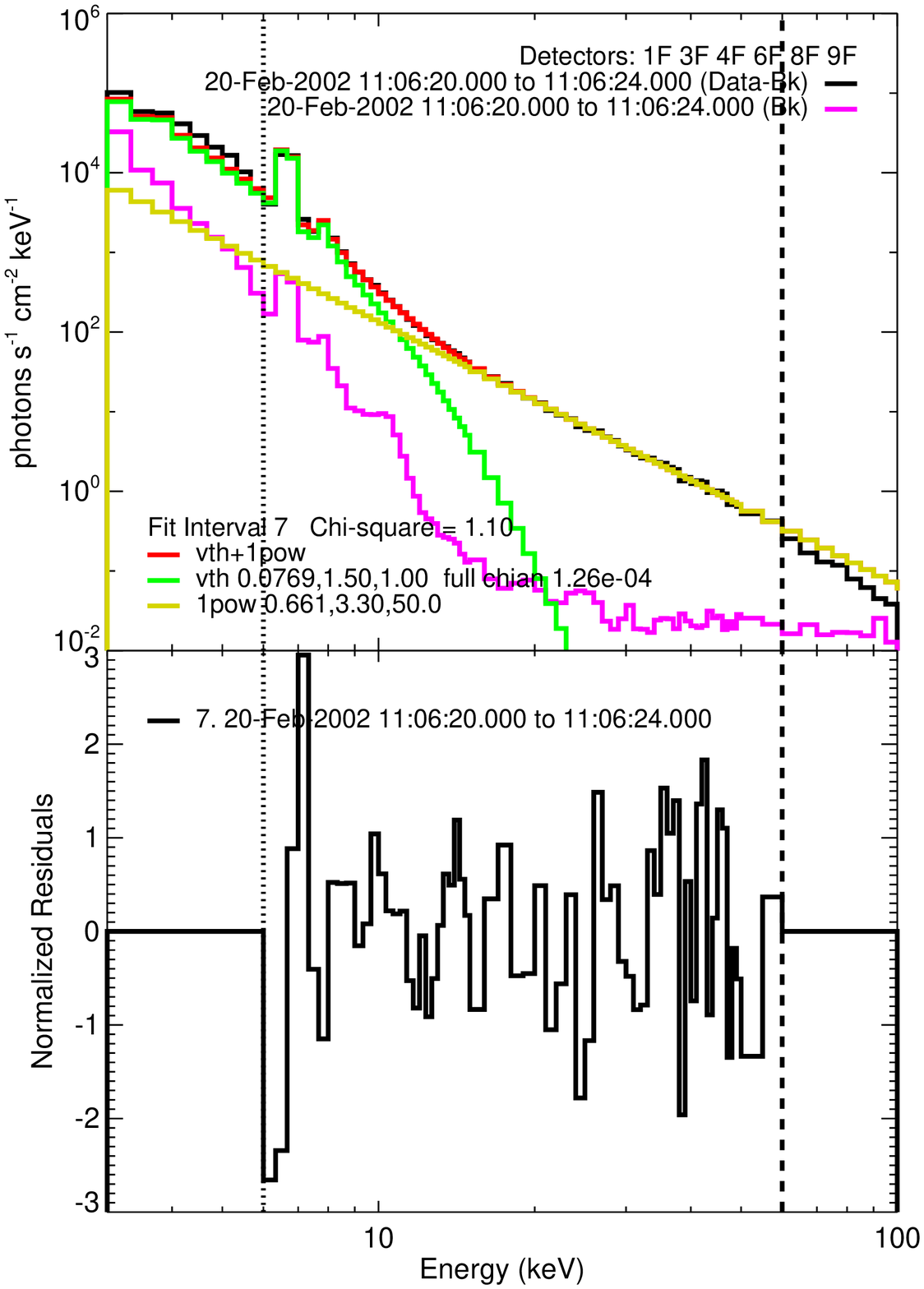}}
\caption{Top panel: photon flux spectrum for the peak time interval:
11:06:20 to 11:06:24. The magenta line shows the background
and the black line is the photon data with background subtracted. The spectrum of data from 6 to 60 keV is fitted with an isothermal plus a single power-law model.
The green line represents the thermal model and the yellow line is
for the power-law model. The red line is for the total thermal plus
power-law spectrum. Bottom panel: normalized residuals of the
spectral fit. } \label{fig:2022003ph7_res}
\end{figure}

The semi-calibrated photon flux, which is simple and fast to obtain,
may give a sufficient approximation of the photon flux from the
source at high photon energies, but the non-diagonal elements of the
response matrix become important at low energies. By carrying out a
full spectral fit, which is much more time-consuming, and
determining the model parameters from this, we can check for
consistency with the results based on the semi-calibrated flux. With
the spectral fitting package OSPEX, the spectrum of counts from 6 to
60 keV are fitted with an isothermal plus a single power-law model
for time bins of 4 seconds from 11:05:56 to 11:06:44 UT.
Note that the model of CHIANTI (5.2) rather than CHIANTI (6.0.1) used in GOES data is applied here due to the lack of implementation of the newer model in OSPEX.
Due to the change of background fluxes in both time and energy, the background
is separately selected for five different energy bands(3 to 6 keV, 6
to 12 keV, 12 to 25 keV, 25 to 50 keV and 50 to 100 keV) both before
and after the flare and fitted with the first order polynomial.
\footnote{The time intervals chosen for fitting the background fluxes are 11:01:52 - 11:02:56 and 11:21:00 - 11:21:56 for 3 - 6 keV, 11:02:00 - 11:02:56 and 11:19:56 - 11:21:00 for 6 - 12 keV, 11:01:56 - 11:02:32 and 11:18:04 - 11:19:00 for 12 - 25 keV, 11:04:08 - 11:04:24 and 11:11:12 - 11:13:16 for 25 - 50 keV, and 11:04:20 - 11:04:48 and 11:08:04 - 11:09:04 for 50 - 100 keV.
We also modeled the background fluxes for eleven energy bins from 3 to 100 keV using a third order polynomial fit with six intervals chosen around the peak time and obtained very similar spectral results.}
Since we are mostly interested in a relatively low energy range, where the
transition between the high and low energy component occurs, an
upper energy bound of 60 keV is chosen to avoid potential spectral steepening at even higher energies \citep{sui2002feb20} and to
ensure adequate counts above background throughout the period of
interest. \citet{asch2002chromos} also fitted the spectra of
this flare with a single power-law from 15 to 50 keV.

We first fit with the temperature, emission measure, power-law index
and normalization of the power-law component as free parameters. The
results are indicated by the black lines in Figure
\ref{fig:2022003para_spec}. Both the emission measure and power-law
index show significant variation during the HXR pulse. The variation
of the temperature, however, is relatively small, between 1.3 and
1.7~keV. To facilitate comparison with the theoretical model, we
then fix the temperature at a typical value of 1.5 keV and do the
spectral fit again. The results are indicated by the red lines in
Figure~\ref{fig:2022003para_spec}. From the $\chi^2$ of the bottom
right panel, we conclude that this model gives a fit to the
observations which is as good as that in the model having the
temperature as a free parameter. The smaller number of free
parameters also gives smaller uncertainties for other model
parameters, and the emission measure (top-left panel) has a smoother
evolution when the temperature is fixed. The power-law spectral
indices (bottom-left panel) from both fits are almost identical and
both have a soft-hard-soft evolution. The goodness of the spectral
fitting is evaluated by the reduced ${\chi}^{2}$ which is shown in
the bottom-right panel. The ${\chi}^{2}$ for both fits are very
similar except during one rise time bin (11:06:04 to 11:06:08) and
one decay time bin (11:06:32 to 11:06:36).

For the sake of simplicity, in the following we only use results
obtained with the model temperature fixed at 1.5 keV. Figure
~\ref{fig:2022003ph7_res} shows the photon spectral fit with the
thermal plus power-law model for the peak time bin (11:06:20 to
11:06:24). The emission measure is $7.69 \pm 0.18\times 10^{47}
\rm{cm^{-3}}$. The normalization of the power-law component at 50
keV is $0.66\pm 0.01$ photons $\rm{s^{-1}cm^{-2}keV^{-1}}$. The
power-law spectral index is 3.3 $\pm$ 0.02. The normalized residuals
are shown below the spectrum. The residuals are between -3 and 3
with slightly larger values at 6-7 keV where the iron emission
lines locate. To model this feature correctly, one needs to fit spectra from individual detectors and take into account small gain changes and pulse pile up. We focus on the transition between thermal and nonthermal components here and leave this caveat for a future investigation.

\begin{figure}
\centerline{\includegraphics[width=0.8\textwidth]{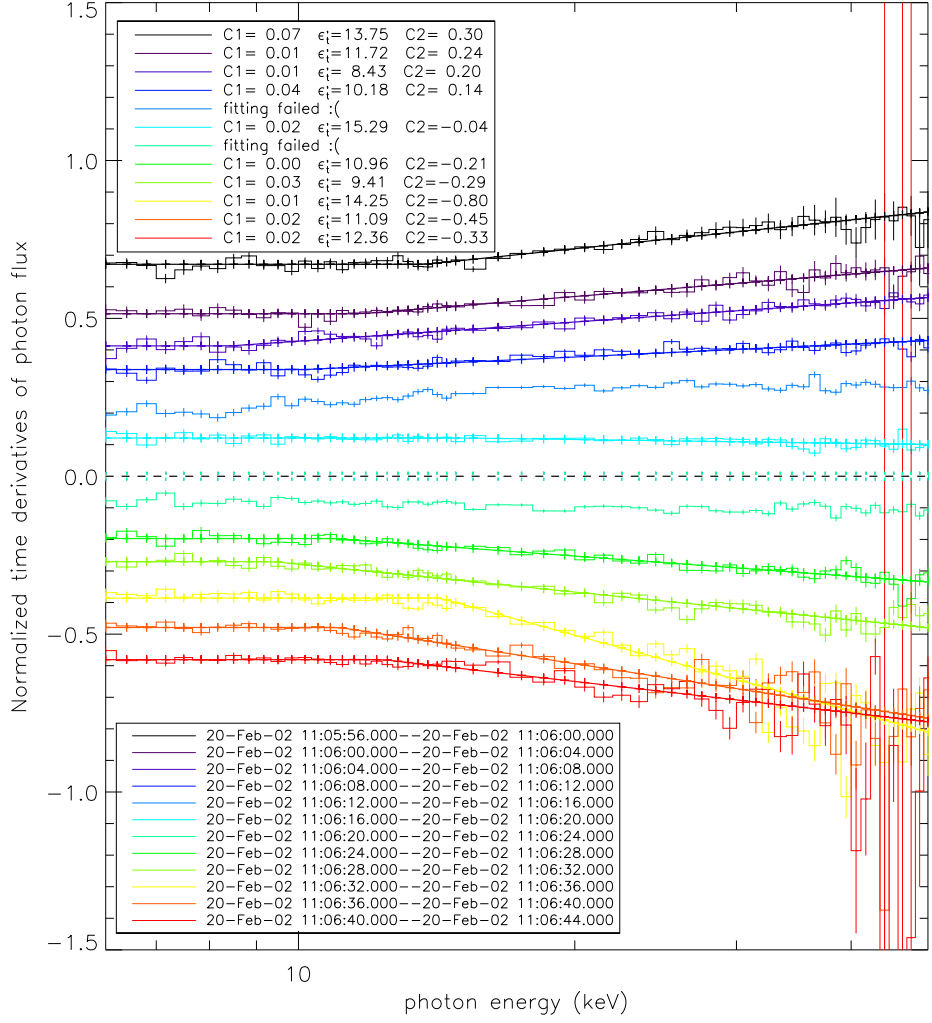}}
\caption{Same as Figure \ref{fig:2022003semi_drv_e} ($R(\epsilon,t)$ in $\rm{s^{-1}}$) but for the photon flux derived from spectral modelling.}
\label{fig:2022003model_DRV_E}
\end{figure}

With the forward-fitted photon fluxes $f$ obtained above, we
calculated the normalized rate of change $R(\epsilon,t) = (d
f(\epsilon,t)/dt)/f(\epsilon,t)$ with the same regularized method as
was used on the semi-calibrated photon fluxes.
Figure~\ref{fig:2022003model_DRV_E} shows the energy dependence of
the rate of change with different colors representing different time
bins. The errors are large for high energies because the photon
fluxes at high energies are low and their relative errors are big.
The solid lines are the model fit shown in
Eq.~(\ref{equ:DRV_E_fit}). Notice that the 5th and 7th lines are
again not fitted with the model.

\begin{figure}
\centerline{\includegraphics[width=0.8\textwidth]{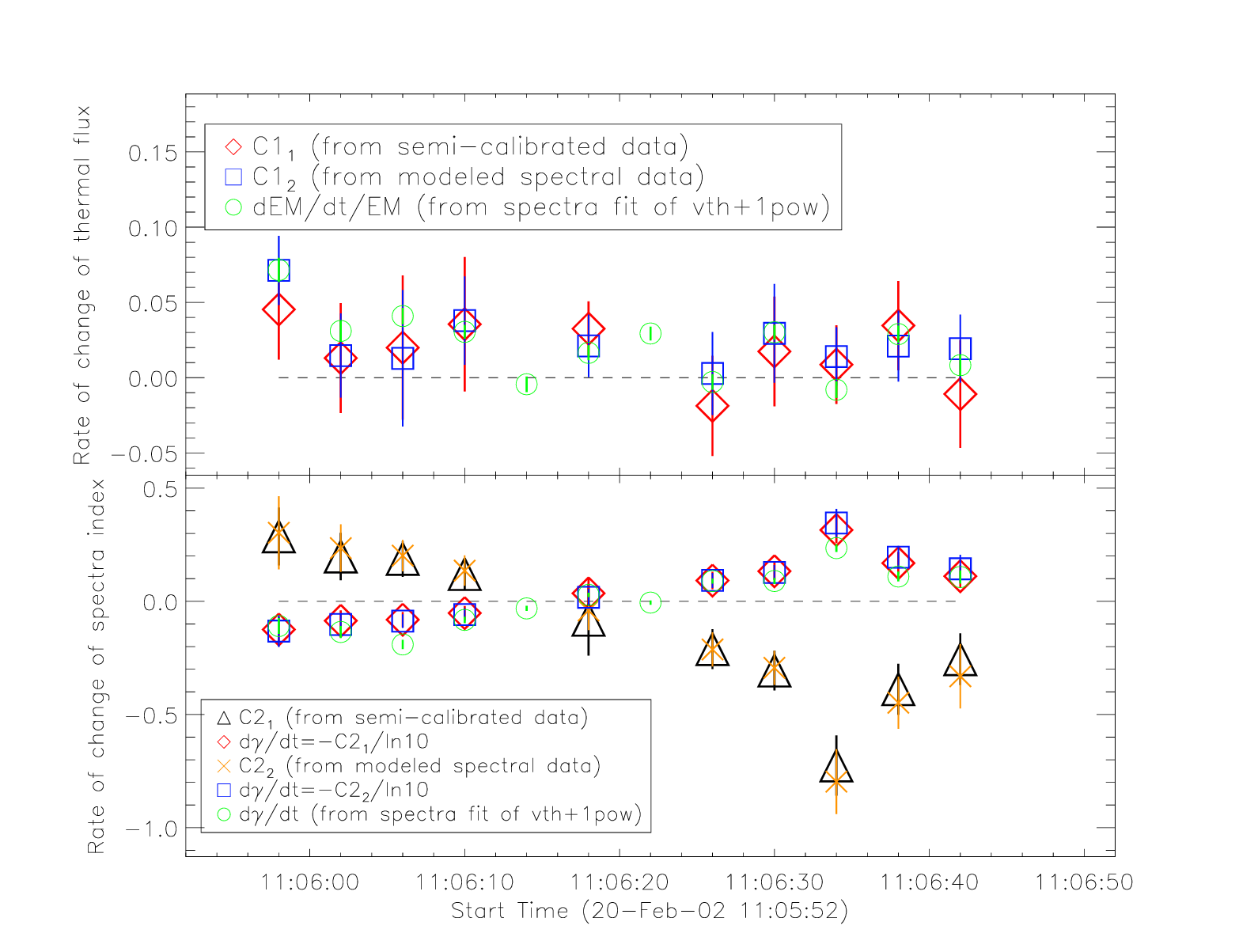}}
\caption{\emph{Top panel}: comparison of the rate of change of
low-energy photon flux $C1_1$ (red diamonds) derived with the semi-calibrated data, $C1_2$ (blue squares) with the modelled spectral data, and the rate of change of the emission measure $\dot{EM}/{EM}$
(green circles). \emph{Bottom panel}: comparison of parameters $C2_1$ (black triangles) and $C2_2$ (orange crosses) derived with the semi-calibrated data and modelled spectral data respectively.
The rate of change of the spectral index $\dot{\gamma}$ can be obtained from $C2_1$ and $C2_2$ as shown in the legend. Red diamonds, blue squares and green circles represent $\dot{\gamma}$ from semi-calibrated photon fluxes, from modelled spectral photon fluxes, and from the time derivatives of the power-law index respectively. Error bars of $\dot{EM}/{EM}$ and $d{\gamma}/dt$ (both in green circles) are obtained with the regularized method. Error bars of all the other parameters indicate the $1 \sigma$ uncertainties of curve fit in Figures~\ref{fig:2022003semi_drv_e} and~\ref{fig:2022003model_DRV_E} and described by Eq.~(\ref{equ:DRV_E_fit}).}
\label{fig:2022003_C1GM}
\end{figure}

According to Eqs.~\ref{equ:C1} and~\ref{equ:C2}, the rate of change of the $EM(t)$ and $\gamma(t)$ can be obtained
directly from the semi-calibrated photon flux, and from the spectral
fits with the regularized method for derivatives
\citep{kontar2005regularized}. Figure ~\ref{fig:2022003_C1GM}
compares the rate of change of the emission measure and the
power-law spectral index with the above three methods: first with
the semi-calibrated data (see Figure~\ref{fig:2022003semi_drv_e}), second with the photon flux derived from spectral fit (see Figure~\ref{fig:2022003model_DRV_E}), and third directly from the evolution of these parameters determined in the spectral fit (see Figure~\ref{fig:2022003para_spec}).
One can see that these
rates obtained with different methods are consistent. The rate of
change of the emission measure is positive in most of the time bins.
This indicates that the thermal emission is increasing nearly
monotonically during the HXR pulse. The time evolution of
$\dot{\gamma}$ is also consistent with the soft-hard-soft evolution.
The regularized method provides a powerful means to quantify this
behavior.

\subsection{Transition and Pivot energy}
\label{s:pivot}

\begin{figure}
\centerline{\includegraphics[width=0.8\textwidth]{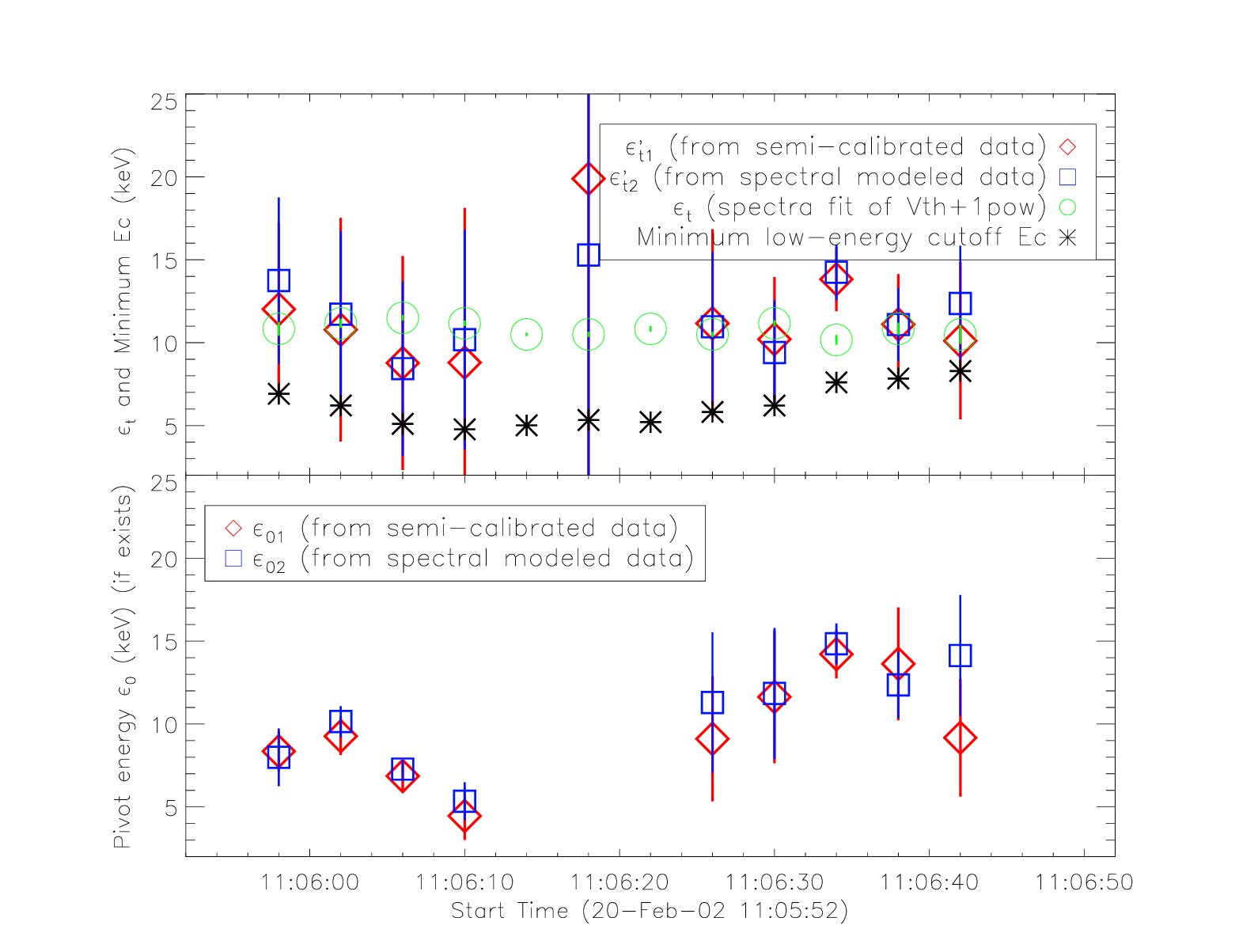}}
\caption{\emph{Top panel}: evolution of the transition energy between the low and high energy component derived from variability of the semi-calibrated data (red diamonds), the modelled spectral data (blue squares) and from the isothermal plus power-law spectral fit (green circles). The error bars of $\epsilon^\prime_{t1}$ and $\epsilon^\prime_{t2}$ show, respectively, the uncertainties of curve fit in Figures~\ref{fig:2022003semi_drv_e} and~\ref{fig:2022003model_DRV_E} as described by Eq.~(\ref{equ:DRV_E_fit}). The error of $\epsilon_t$ is the standard deviation of 5000 simulated intersections between thermal and nonthermal components taking into account uncertainties of all the spectral fitting parameters. Also shown is the possible minimum
electron low-energy cutoff $E_c$ (Section~\ref{s:discussion}).
\emph{Bottom panel}: pivot energy $\epsilon_0$ with $1 \sigma$ errors (see Section~\ref{s:pivot} for more details) derived from the variability of the semi-calibrated data (red diamonds) and modelled spectral photon fluxes (blue squares).   }
\label{fig:2022003_EtEp}
\end{figure}

With the results obtained above, one can check the consistency of
the model given by Equation (\ref{equ:model}) and test whether there
is one pivot energy $\epsilon_0$ for the duration of the whole HXR
pulse. The transition energy between the two temporal components
$\epsilon^\prime_t$ and between the two spectral components
$\epsilon_t$ should be comparable if the two spectral components
have distinct energy and time dependences, as expected.
For results obtained with
the spectral fitting in Section \ref{s:spectral_fit}, good spectral
fits will guarantee that $\epsilon^\prime_t$ be comparable to
$\epsilon_t$. However, with the semi-calibrated data, these two
transition energies characterized the spectral evolution in two
distinct dimensions. There is no guarantee that the two spectral
components will match the two temporal components self-consistently.
The top panel of Figure \ref{fig:2022003_EtEp} shows $\epsilon_t$
and $\epsilon^\prime_t$ obtained from the analyses above.
$\epsilon_t$ has a relative error of a few percent and varies between 10 and 12 keV, which is consistent with a constant value of $\sim 11$ keV. Although $\epsilon^\prime_t$ varies in a larger energy range of 8 to 20 keV, the relative errors are greater than $15\%$ and its values are also consistent with a constant of $\sim 11$ keV. The
agreement of these quantities implies consistency of the model.
The big error bars of $\epsilon^\prime_t$ are due to the uncertainty of determining the cross point of the two lines in the fitting shown in Figures~\ref{fig:2022003semi_drv_e} and~\ref{fig:2022003model_DRV_E}. This may reveal a complicated physical process where the transition between the slow-varying gradual component and the impulsive component is rather an energy range than a single point.

The usual method of determining $\epsilon_0$ by spectroscopic
fitting depends on the assumed spectral model.
We instead use the above rate of change study to derive $\epsilon_0$ and
its errors for each interval. From equation~(\ref{equ:DRV_E_fit}),
one can show that the rate of change of the photon flux of the
power-law component is zero at
\begin{equation}\label{equ:E0}
\epsilon_0=\epsilon^\prime_t \cdot {10^{-C1/C2}}\,.
\end{equation}
This is the pivot energy at a given time interval.  With parameters
$C1$, $C2$, and $\epsilon^\prime_t$ obtained above, we calculated the pivot
energy $\epsilon_0$ as shown in the bottom panel of Figure
~\ref{fig:2022003_EtEp}.
The error of $\epsilon_0$ is taken as the standard deviation of simulated $\epsilon_0$ with 5000 sampling points of $\epsilon^\prime_t$, $C1$, and $C2$ distributed within their respective $1 \sigma$ range of a Gaussian distribution. The results are consistent with a constant value of $\sim 9$ keV except for the fourth and tenth intervals where deviations of $\epsilon_0$ from $9$ keV are greater than $3 \sigma$. These values of the pivot energy are also less than those obtained by \citet{battaglia2006} for the loop top source of a few other flares, but are in agreement with these of the footpoint sources especially for values in the HXR decay phase. Since $C1$ is mostly positive and $C2$ evolves from positive to negative values from the HXR rise to decay phase, $10^{-C1/C2}$ evolves from less than 1 to greater than 1, implying that $\epsilon_0$ increases from $\lesssim \epsilon^\prime_t$ to $\gtrsim \epsilon^\prime_t$. It should be noticed that when $C2$ approaches zero, the amplitude of $C1/C2$ approaches infinity and $10^{-C1/C2}$ approaches either $0$ or infinity, both of which are not physical. It also leads to huge error bars for time bins near and after the HXR peak. Indeed, a sampling of $C2$ over a full Gaussian distribution after the peak (when $C2<0$ and $C1>0$) will lead to infinite values of $\epsilon_0$, whose distribution is poorly fitted with a Gaussian. Theoretically, for HXR pulses with soft-hard-soft spectral evolution, if the transition energy and temperature do not change significantly and the emission measure has a gradual and monotonic increase, the pivot energy in the
decay phase should be higher than that in the rise phase, as indicated in Figure \ref{fig:2022003_EtEp}.
This result is in agreement with previous
studies \citep{grigis05, battaglia2006} and may be attributed to an effect of chromospheric evaporation \citep{liu2010elementary}.
The pivot energy $\epsilon_0$ is also
comparable to the transition energies $\epsilon_t$ and
$\epsilon^\prime_t$, in agreement with the scenario where energetic
electrons are accelerated from a low-energy thermal background
plasma \citep{benz1977, petrosian_liu2004, grigis2006electron}.

\section{Discussion}
\label{s:discussion}

We started with the hypothesis that there are two distinct emission
components with low energy photons evolving gradually and high
energy photons having a rapid evolution. One consequence of this is
that, as long as the temperature of the thermal component varies
slowly (much slower than the emission measure), which simplifies the model significantly [see eq. (\ref{equ:ratet})],  the break energy between non-thermal and thermal emission in
the photon spectrum should be comparable to the transition energy
between slowly- and rapidly-varying photon fluxes found by
evaluating time derivatives. Within the uncertainties of this
method, we have demonstrated that this is the case (top panel of
Figure~\ref{fig:2022003_EtEp}), and that the transition energies are
always around $11$~keV. However, it is clear that there are
substantial error bars on the values of the transition energy
$\epsilon_t^\prime$, which are relatively independent of whether the
spectral fitting approach (model-dependent) or the semi-calibrated
approach (model independent) is used. In fact it is not possible to
pin down the gradual/rapid boundary within about $\pm 5$~keV
throughout most of the flare, especially near the HXR peak. This is
due to the difficulty of determining the folding point of the broken
line. Better data with much higher count rates and lower statistical
errors are required --- for example, a more intense but equally
simple flare --- to examine whether or not such a boundary can be
more clearly identified.

To understand further the relationship between the high and low
energy emission components produced presumably by two distinct
electron populations through the bremsstrahlung process, we
investigate the electron numbers in each population.
In one version of the standard model, it is postulated that electrons are accelerated at a reconnection current sheet and the acceleration process is decoupled from the electron transport and magnetic field evolution after the reconnection \citep{asch1998deconvolution}, which would imply an ideal Neupert effect not wholly supported by observations \citep{veronig2005}. This scenario also encounters the well-known number problem \citep{fletcher2008}. Given the high energy release during some large flares \citep{emslie2004, emslie2005}, energy flows likely play more important roles than nonthermal electron fluxes in our exploration of the physics in the impulsive phase. It is possible that a significant fraction of the magnetic energy is converted into particle energies after the reconnection during the relaxation of magnetic field lines. The reconnection only permits the changes of magnetic field topology and may not correspond to the dominant energy dissipation and particle acceleration process, which can proceed after the reconnection \citep{fletcher2008}.
In the context of stochastic particle acceleration, it is usual to
assume that electrons arriving at the chromosphere are accelerated
out of a population in the loop \citep{petrosian_liu2004}, therefore
--- assuming for simplicity no magnetic convergence --- the number
density of non-thermal electrons should be no larger than the loop
number density. \citet{sui2002feb20} suggested that high-energy electrons might be accelerated from a cool background plasma not observed in X-rays. There is no observational evidence for such a cold background. Theoretically, it is also difficult to understand why the acceleration should proceed in relatively cool regions given the microphysics of particle energization by electric fields is the same for both thermal and nonthermal populations. Moreover, as we will show below, the observed thermal plasma is dense enough to provide electrons responsible for the high energy emission. The assumption of a cooler background source for the high energy electrons appears to be unnecessary. In the following, we will assume that electrons producing the high energy component are accelerated from the observed thermal component.
The number density of a nonthermal electron beam can
be estimated as
\begin{equation}\label{equ:n_nth}
n_{nth}= {P_c \over { \bar{E} \bar{v_e}  A_{HXR}} },
\end{equation}
where $P_c$ is the power in electrons of energy greater than $E_c$,
$A_{HXR}$ is the area of HXR footpoints where electrons enter the
chromosphere and can be estimated from flare images, $\bar{E}$ is
the average electron energy and $\bar{v_e}$ is the corresponding
electron velocity. $P_c$ is given by
\begin{equation}\label{equ:power2}
P_c={A_E \over {\delta -2}} (E_c/{\rm keV})^{-(\delta -2)}  [\rm{keV
s^{-1}}],
\end{equation} where $\delta$ is the spectral index of the underlying non-thermal electron flux spectrum ($\delta = \gamma +1$ in the collisional thick target model),
$A_E$ is a normalization parameter and is numerically related to the
photon spectrum \citep{brown1971,saint2005, fletcher2007trace}:
\begin{equation}\label{equ:A_E}
A_E=6.44 \times 10^{33} {{\gamma (\gamma-1)} \over B(\gamma-1, 1/2)}
A_\epsilon ,
\end{equation}
where $B$ is the beta function, and $A_\epsilon$ is  the
normalization of the power-law fit to the photon spectrum
$f(\epsilon)=A_\epsilon (\epsilon/{\rm keV})^{-\gamma}$ (in photons
$\rm{s^{-1} cm^{-2} keV^{-1}}$). It can be shown that $\bar{E}\equiv
\int_{E_c}^{\infty} F(E) E d E/\int_{E_c} F(E) d E =(\delta-1)E_c/(\delta-2)$
where $F(E)=A_E (E/{\rm keV})^{-\delta}$ is the electron flux injected into the footpoints
\citep{brown1971}. $\bar{v_e} \equiv \int_{E_c}^{\infty} F(E) d E/\int_{E_c}^{\infty}
F(E)/v_e(E) d E = (\delta-1/2)v_{e}(E_c)/(\delta-1)$, where we have
assumed the nonthermal electron distribution is given by
$F(E)/(v_e(E)A_{HXR})$.

%$V_{SXR}$ can be evaluated by assuming that the source is a loop
%whose length corresponds to the SXR source length $L_{SXR}$ and
%whose cross-sectional area is equal to half of the footpoint area
%$A_{HXR}/2$ for a double footpoint flare.
The thermal electron number density is estimated as
$n_{th}=(EM/V_{SXR})^{1/2}$, where $EM$ is the emission measure of
the thermal component and $V_{SXR}$ is the volume of the SXR thermal
coronal loop. For the sake of simplicity, $V_{SXR}$ is evaluated
as $A_{SXR}^{3/2}$ where $A_{SXR}$ is the projected area of the
observed thermal coronal source. The areas within the 30\% contours
of the maximum value of 25-100 keV (for HXR source) and 6-9 keV (for
SXR source) images can be obtained from the Pixon images directly.
%The area of the SXR source is then equal to
%$L_{SXR}(2A_{HXR}/\pi)^{1/2}$.
From the peak-time 20-second integrated image shown in Figure
\ref{fig:image2}, we estimate $A_{HXR}$ to be 20 square arcsec,
$A_{SXR}$ to be 100 square arcsec, and $V_{SXR}$ is then about 1000
cubic arcsec. The corresponding $n_{th}$ is greater than $\sim 10^{11}$ cm$^{-3}$.

%\begin{figure}
%\centerline{\includegraphics[width=0.8\textwidth]{f13}}
%\caption{The minimum low-energy cutoff to the electron spectrum
%$E_c$ (keV) in Eq. (\ref{equ:Ec}) as a function of thermal emission
%measure $EM$ ($10^{47} \rm{cm^{-3}}$) and photon spectral index
%$\gamma$, both obtained from photon spectral fitting in Section
%\ref{s:spectral_fit}. The time sequence can be traced by noting that
%the EM is increasing nearly monotonically during the flare peak as
%shown in Figure \ref{fig:2022003para_spec}. }
%\label{fig:2022003_Ec_im}
%\end{figure}

%\begin{figure}[]
%\begin{center}
%\includegraphics[width=0.45\textwidth]{f14a}
%\includegraphics[width=0.45\textwidth]{f14b}
%\end{center}
%\caption{The minimum low-energy cutoff $E_c$ (keV) as a function of
%the size of SXR source $A_{SXR}$ (arcsec$^2$) and size of HXR
%footpoints $A_{HXR}$ (arcsec$^2$). $A_{SXR}$ and $A_{HXR}$ change
%between 10 to 200 and 10 to 40 arcsec$^2$ respectively. \emph{Left
%panel}: $\gamma=3.3$ and $EM=7.7 \times 10^{47} \rm{cm^{-3}}$ as
%parameters for the calculation. \emph{Right panel}: $\gamma=5.8$ and
%$EM=9.8 \times 10^{47} \rm{cm^{-3}}$.} \label{fig:2022003Ec}
%\end{figure}

The fraction of the thermal electrons accelerated into a non-thermal
distribution $\alpha$ should be less than 1. We then have
$n_{nth}\le n_{th}$ and
\begin{equation}\label{equ:Ec}
\left({E_c\over {\rm keV}}\right)^{-\delta +1/2}  = \alpha \left({2
EM \over m_e }\right)^{1/2} {{\delta-1/2} \over A_E } {A_{HXR} \over
{V_{SXR}}^{1/2} }.
\end{equation}
Setting $\alpha = 1$, which would correspond to the minimum possible
value of $E_c$, and with the parameters $EM$, $\gamma$, $A_\epsilon$
from Section \ref{s:spectral_fit}, we calculate the low limit of
$E_c$ in each 4 second time bin of the spectral fitting.
%and the dependence of $E_c$ on $EM$ and $\gamma$ is shown in Figure~
%\ref{fig:2022003_Ec_im}. It can be seen that larger spectral index
%results in an increase of the minimum $E_c$. According to
%Equation~\ref{equ:Ec}, the increase of $EM$ will lead to the
%decrease of the minimum $E_c$. However, this effect is not as
%obvious as that from the changes of $\gamma$ as shown in Figure~
%\ref{fig:2022003_Ec_im}.
A comparison of low limit of $E_c$ and the
transition energies derived from variability of the semi-calibrated
data ($\epsilon^\prime_{t1}$), the modelled spectral data
($\epsilon^\prime_{t2}$) and from the isothermal plus power-law
spectral fit ($\epsilon_t$) is presented in
Figure~\ref{fig:2022003_EtEp}. At the peak of the flare, the minimum
possible $E_c$ is around 5.2~keV, and throughout the event it is
always smaller than the transition energy $\epsilon_t$, which is
around 11 keV. This is fully consistent with the stochastic
acceleration model where non-thermal electrons are accelerated from
a thermal background \citep{benz1977, petrosian_liu2004,
grigis2006electron}. We note that $\delta$ is always greater than 4 for this flare, the $E_c$ obtained with equation (\ref{equ:Ec}) is rather insensitive to the poorly-determined source size $A_{HXR}$ and $A_{SXR}$.

%we evaluate $A_{HXR}$ and $V_{SXR}$
%with a simple single-loop model so that $A_{HXR} = \pi {D_{HXR}}^{2}
%/2$ and $V_{SXR}= L_{SXR} A_{HXR}$, where $D_{HXR}$ is the diameter
%of the HXR footpoints and $L_{SXR}$ is the length of the SXR loop.

%To accommodate reasonable uncertainties in the sizes of the HXR
%footpoints and SXR loop sources obtained from RHESSI imaging, we
%evaluate the values of $A_{HXR}$ and $A_{SXR}$ within reasonable
%limits. We show $E_c$ versus different values of $A_{HXR}$ (changing
%from 10 to 40 arcsec$^2$) and $A_{SXR}$ (changing from 10 to 200
%arcsec$^2$) in Figure \ref{fig:2022003Ec}. $E_c$ shown in the left
%panel is calculated from $\gamma$ and $EM$ obtained from spectral
%fitting of the peak of flare from 11:06:20 to 11:06:24 shown in
%Figure \ref{fig:2022003ph7_res}. $E_c$ of the right panel employs
%parameters from the spectral fitting of the time bin of 11:06:40 to
%11:06:44. It is clear that with larger SXR loop and smaller
%footpoint size the low limit of $E_c$ is increased. However it is
%not sensitive to X-ray source sizes $A_{SXR}$ and $A_{HXR}$, which
%are not well determined from observations.

Rearranging Equation (\ref{equ:Ec}), and keeping other factors
constant, $\alpha$ and $E_c$ vary as $\alpha \propto
E_c^{-(\delta-1/2)}$. If we set the low-energy cutoff $E_c \sim
\epsilon_t \sim$ 10~keV, with the low limit of $E_c = 5$ keV at
$\alpha = 1$ for the flare peak where
$\delta=\gamma+1=4.3$, this gives $\alpha = 0.07$. In other words,
accelerating 7\% of the hot thermal distribution in the loop would
satisfy the footpoint requirements. One may further assume that these accelerated
electrons come from the high energy tail of the thermal distribution. In the
tail of a Maxwellian, the fraction $\alpha_\chi$ of electrons with
energy above $E = \chi k T$ is $\alpha_\chi =
\sqrt{4\chi/\pi}\exp(-\chi)$, so $\alpha = 0.07$ corresponds to
$\chi = 3.4$. Since $kT = 1.5$~keV, the non-thermal population would
correspond to the accelerated tail of electrons with initial energy
above 5.1~keV. This in turn sets a requirement that the acceleration
timescale would have to be less than the electron-electron Coulomb
collision timescale $\tau_{ee}$ for an electron of energy 5.1~keV
in a Maxwellian plasma of temperature 1.5~keV and density $4.5\times
10^{10}\rm{cm}^{-3}$ (using $EM, V_{SXR}$ determined above). The
value of $\tau_{ee}$ for an electron in the core of such a
distribution is 0.02~s, and that for an electron of energy $\chi kT$
is approximately $\chi^{3/2}\tau_{ee}$ or 0.13~s in this case.
Note that this threshold energy should not be compared with the transition energies in the overall photon spectra directly. The electron transport and X-ray emission processes will make the transition energy in the electron population different from that of the emitted photons. In the context of stochastic acceleration in the flare loop, a higher break energy in the overall photon spectrum than in the electron distribution in the corona acceleration site implies that the electron escape timescale from the acceleration site to the footpoints is relatively long compared to the Coulomb collision timescale at these break energies so that the HXR fluxes from the footpoints are suppressed \citep{petrosian_liu2004}.

It should here be remarked that the low values obtained for $E_c$
call into question the application of the cold collisional thick
target. \citet{emslie2003} shows that the cold collisional
thick target loss rates are overestimated for electrons of energy
less than $5\ kT$. Though we have put bounds on $E_c$, it remains a
fit parameter, and one can instead consider the energetics of an
injected non-thermal electron distribution extending from $kT$,
which merges into the ambient thermal distribution. For this event
at its peak ($\gamma = 3.3$) the collisional thick target power
requirement above $5kT$ is $7.3\times 10^{28}{\rm erg~s^{-1}}$, which is already too high compared with the total radiation energy of $\sim 10^{29}$ ergs obtained from the GOES observation.
Following the calculation of \citet{emslie2003} gives about
35 times as much as this in total injected power above $kT$, which further demonstrates the necessity of going beyond the classical cold thick target model.

The discussion above assumes a beam of electrons with a power-law
distribution. Considering the pitch-angle scattering, the average
electron velocity will be lower giving rise to a higher
local non-thermal electron density and hence a larger $E_c$.
Indeed, X-ray images reveal more complex structure near the
transition energy between thermal and non-thermal components.
The separation of the emission into two distinct components
is rather ambiguous near the transition energy. It is possible
that this separation is an artifact of a simplified model
of X-ray spectrum and the energy dependence of the rate
of change of the photon fluxes. Two distinct electron populations
can be due to different physical processes, which dominate
at different energies, with the electron behavior varying
considerably from low to high energies and the apparent distinction
between the low and high energy emission components being
just a consequence of these processes.

\section{Conclusions}
\label{s:conclusions}

We have developed a new method to study in detail the temporal
evolution of thermal and non-thermal photon fluxes in solar flares.
The application of this method to a flare on Feburary 20 2002
demonstrates that as expected, the low energy part of the spectrum
evolves slowly, and the high energy part evolves rapidly, with an
intermediate range between a few keV and 20 keV where the behavior
is in transition. The data support the scenario in which the
non-thermal component of the flare spectrum is impulsive, and the
thermal component is gradual, in that the transition energies
between these two behaviors are the same within errors whether
examined in time or in energy.

However, although in the spectral fitting exercise it is possible to
make a clean separation between a non-thermal, impulsive component,
and a thermal, gradual component, time evolution gives a more
ambiguous picture, due to the large error bars. Imaging is also
ambiguous, with no clear distinction between footpoints and loops in
the energy range around 9-25~keV.  Therefore we must leave open the
possibility that the electrons form a continuous distribution over
this range. Further studies with larger flares should help to
improve the precision with which we can identify the transition
between gradual and impulsive behavior.

Finally, the presence of a single pivot point throughout the flare
is not supported by our analysis, though a pivot `range' is. There
is some evidence of a slightly higher value for this pivot
range in the decay phase than in the rise phase.

\acknowledgements The authors are very grateful to the referee for helpful comments and we would like to thank Kim Tolbert, Hugh
Hudson, Marina Battaglia and Iain Hannah for useful RHESSI
techniques and fruitful discussions. Jingnan Guo would like to thank
Eckart Marsch, J\"{o}rg B\"{u}chner and Weiqun Gan for sharing their
substantial knowledge of flares, plasma physics and particle
accelerations. The work is supported by the European Community
Research Training Network project SOLAIRE (MTRN-CT-2006-035484),
STFC Rolling Grant ST / F002637 /1, Leverhulme Foundation Grant
F00-179A, NNSFC (10833007), 973 (20110B811402), and STFC Advanced Fellowship.

\clearpage

\bibliographystyle{apj}
\bibliography{Drv_E_bib} % your references Yourfile.bib

\begin{thebibliography}{46}
\expandafter\ifx\csname natexlab\endcsname\relax\def\natexlab#1{#1}\fi

\bibitem[{{Aschwanden}(1998)}]{asch1998deconvolution}
{Aschwanden}, M.~J. 1998, \apj, 502, 455

\bibitem[{{Aschwanden}(2002)}]{asch2002paa}
---. 2002, Space~Sci.~Rev., 101, 1

\bibitem[{{Aschwanden}(2005)}]{asch2005psc}
---. 2005, {Physics of the Solar Corona. An Introduction with Problems and
  Solutions (2nd edition)}

\bibitem[{{Aschwanden} {et~al.}(2002){Aschwanden}, {Brown}, \&
  {Kontar}}]{asch2002chromos}
{Aschwanden}, M.~J., {Brown}, J.~C., \& {Kontar}, E.~P. 2002, \solphys, 210,
  383

\bibitem[{{Aschwanden} {et~al.}(1996{\natexlab{a}}){Aschwanden}, {Kosugi},
  {Hudson}, {Wills}, \& {Schwartz}}]{asch1996scaling}
{Aschwanden}, M.~J., {Kosugi}, T., {Hudson}, H.~S., {Wills}, M.~J., \&
  {Schwartz}, R.~A. 1996{\natexlab{a}}, \apj, 470, 1198

\bibitem[{{Aschwanden} {et~al.}(1996{\natexlab{b}}){Aschwanden}, {Wills},
  {Hudson}, {Kosugi}, \& {Schwartz}}]{asch1996loop}
{Aschwanden}, M.~J., {Wills}, M.~J., {Hudson}, H.~S., {Kosugi}, T., \&
  {Schwartz}, R.~A. 1996{\natexlab{b}}, \apj, 468, 398

\bibitem[{{Bai} \& {Ramaty}(1979)}]{bai1979}
{Bai}, T., \& {Ramaty}, R. 1979, \apj, 227, 1072

\bibitem[{Battaglia \& Benz(2006)}]{battaglia2006}
Battaglia, M., \& Benz, A. 2006, Astronomy and astrophysics, 456, 751

\bibitem[{{Battaglia} {et~al.}(2009){Battaglia}, {Fletcher}, \&
  {Benz}}]{battaglia2009}
{Battaglia}, M., {Fletcher}, L., \& {Benz}, A.~O. 2009, \aap, 498, 891

\bibitem[{{Benz}(1977)}]{benz1977}
{Benz}, A.~O. 1977, \apj, 211, 270

\bibitem[{Brown(1971)}]{brown1971}
Brown, J. 1971, Sol.~Phys., 18, 489

\bibitem[{{Dennis} \& {Zarro}(1993)}]{dennis1993}
{Dennis}, B.~R., \& {Zarro}, D.~M. 1993, \solphys, 146, 177

\bibitem[{{Dere} {et~al.}(2009){Dere}, {Landi}, {Young}, {Del Zanna},
  {Landini}, \& {Mason}}]{dere2009}
{Dere}, K.~P., {Landi}, E., {Young}, P.~R., {Del Zanna}, G., {Landini}, M., \&
  {Mason}, H.~E. 2009, \aap, 498, 915

\bibitem[{{Emslie}(2003)}]{emslie2003}
{Emslie}, A.~G. 2003, \apjl, 595, L119

\bibitem[{{Emslie} {et~al.}(2005){Emslie}, {Dennis}, {Holman}, \&
  {Hudson}}]{emslie2005}
{Emslie}, A.~G., {Dennis}, B.~R., {Holman}, G.~D., \& {Hudson}, H.~S. 2005,
  Journal of Geophysical Research (Space Physics), 110, 11103

\bibitem[{{Emslie} {et~al.}(2004){Emslie}, {Kucharek}, {Dennis}, {Gopalswamy},
  {Holman}, {Share}, {Vourlidas}, {Forbes}, {Gallagher}, {Mason}, {Metcalf},
  {Mewaldt}, {Murphy}, {Schwartz}, \& {Zurbuchen}}]{emslie2004}
{Emslie}, A.~G., {Kucharek}, H., {Dennis}, B.~R., {Gopalswamy}, N., {Holman},
  G.~D., {Share}, G.~H., {Vourlidas}, A., {Forbes}, T.~G., {Gallagher}, P.~T.,
  {Mason}, G.~M., {Metcalf}, T.~R., {Mewaldt}, R.~A., {Murphy}, R.~J.,
  {Schwartz}, R.~A., \& {Zurbuchen}, T.~H. 2004, Journal of Geophysical
  Research (Space Physics), 109, 10104

\bibitem[{{Feldman} {et~al.}(2004){Feldman}, {Dammasch}, {Landi}, \&
  {Doschek}}]{feldman2004}
{Feldman}, U., {Dammasch}, I., {Landi}, E., \& {Doschek}, G.~A. 2004, \apj,
  609, 439

\bibitem[{{Fisher}(1989)}]{fisher1989}
{Fisher}, G.~H. 1989, \apj, 346, 1019

\bibitem[{{Fletcher} {et~al.}(2007){Fletcher}, {Hannah}, {Hudson}, \&
  {Metcalf}}]{fletcher2007trace}
{Fletcher}, L., {Hannah}, I.~G., {Hudson}, H.~S., \& {Metcalf}, T.~R. 2007,
  \apj, 656, 1187

\bibitem[{{Fletcher} \& {Hudson}(2008)}]{fletcher2008}
{Fletcher}, L., \& {Hudson}, H.~S. 2008, \apj, 675, 1645

\bibitem[{{Gallagher} {et~al.}(2002){Gallagher}, {Dennis}, {Krucker},
  {Schwartz}, \& {Tolbert}}]{gallagher2002}
{Gallagher}, P.~T., {Dennis}, B.~R., {Krucker}, S., {Schwartz}, R.~A., \&
  {Tolbert}, A.~K. 2002, \solphys, 210, 341

\bibitem[{Gan(1998)}]{gan1998invariable}
Gan, W. 1998, Astrophysics and Space Science, 260, 515

\bibitem[{{Grigis} \& {Benz}(2004)}]{grigis2004spectral}
{Grigis}, P.~C., \& {Benz}, A.~O. 2004, \aap, 426, 1093

\bibitem[{{Grigis} \& {Benz}(2005)}]{grigis05}
---. 2005, \aap, 434, 1173

\bibitem[{{Grigis} \& {Benz}(2006)}]{grigis2006electron}
---. 2006, A\&A, 458, 641

\bibitem[{{Hannah} {et~al.}(2009){Hannah}, {Kontar}, \&
  {Sirenko}}]{hannah2009wave}
{Hannah}, I.~G., {Kontar}, E.~P., \& {Sirenko}, O.~K. 2009, \apjl, 707, L45

\bibitem[{{Kane} \& {Anderson}(1970)}]{kane1970}
{Kane}, S.~R., \& {Anderson}, K.~A. 1970, \apj, 162, 1003

\bibitem[{Kontar \& MacKinnon(2005)}]{kontar2005regularized}
Kontar, E., \& MacKinnon, A. 2005, Solar Physics, 227, 299

\bibitem[{{Kontar} {et~al.}(2008){Kontar}, {Dickson}, \& {Ka{\v
  s}parov{\'a}}}]{Kontar2008cutoff}
{Kontar}, E.~P., {Dickson}, E., \& {Ka{\v s}parov{\'a}}, J. 2008, \solphys,
  252, 139

\bibitem[{{Krucker} {et~al.}(2008){Krucker}, {Battaglia}, {Cargill},
  {Fletcher}, {Hudson}, {MacKinnon}, {Masuda}, {Sui}, {Tomczak}, {Veronig},
  {Vlahos}, \& {White}}]{krucker2008HXR}
{Krucker}, S., {Battaglia}, M., {Cargill}, P.~J., {Fletcher}, L., {Hudson},
  H.~S., {MacKinnon}, A.~L., {Masuda}, S., {Sui}, L., {Tomczak}, M., {Veronig},
  A.~L., {Vlahos}, L., \& {White}, S.~M. 2008, \aapr, 16, 155

\bibitem[{Lin {et~al.}(2002)Lin, Dennis, Hurford, Smith, Zehnder, Harvey,
  Curtis, Pankow, Turin, Bester, {et~al.}}]{lin2002rhessi}
Lin, R., Dennis, B., Hurford, G., Smith, D., Zehnder, A., Harvey, P., Curtis,
  D., Pankow, D., Turin, P., Bester, M., {et~al.} 2002, Solar Physics, 210, 3

\bibitem[{{Liu} \& {Fletcher}(2009)}]{liu2009elementary}
{Liu}, S., \& {Fletcher}, L. 2009, \apjl, 701, L34

\bibitem[{{Liu} {et~al.}(2010){Liu}, {Han}, \& {Fletcher}}]{liu2010elementary}
{Liu}, S., {Han}, F., \& {Fletcher}, L. 2010, \apj, 709, 58

\bibitem[{{McAteer} {et~al.}(2007){McAteer}, {Young}, {Ireland}, \&
  {Gallagher}}]{mcateer2007bursty}
{McAteer}, R.~T.~J., {Young}, C.~A., {Ireland}, J., \& {Gallagher}, P.~T. 2007,
  \apj, 662, 691

\bibitem[{{McTiernan} {et~al.}(1999){McTiernan}, {Fisher}, \&
  {Li}}]{mctiernan1999}
{McTiernan}, J.~M., {Fisher}, G.~H., \& {Li}, P. 1999, \apj, 514, 472

\bibitem[{{Miller} {et~al.}(1997){Miller}, {Cargill}, {Emslie}, {Holman},
  {Dennis}, {LaRosa}, {Winglee}, {Benka}, \& {Tsuneta}}]{miller1997jgr}
{Miller}, J.~A., {Cargill}, P.~J., {Emslie}, A.~G., {Holman}, G.~D., {Dennis},
  B.~R., {LaRosa}, T.~N., {Winglee}, R.~M., {Benka}, S.~G., \& {Tsuneta}, S.
  1997, \jgr, 102, 14631

\bibitem[{{Mrozek} \& {Tomczak}(2004)}]{mrozek2004}
{Mrozek}, T., \& {Tomczak}, M. 2004, \aap, 415, 377

\bibitem[{{Neupert}(1968)}]{neupert1968}
{Neupert}, W.~M. 1968, \apjl, 153, L59+

\bibitem[{{Petrosian}(1973)}]{petrosian1973}
{Petrosian}, V. 1973, \apj, 186, 291

\bibitem[{{Petrosian} \& {Liu}(2004)}]{petrosian_liu2004}
{Petrosian}, V., \& {Liu}, S. 2004, \apj, 610, 550

\bibitem[{{Pina} \& {Puetter}(1993)}]{Pina1993}
{Pina}, R.~K., \& {Puetter}, R.~C. 1993, \pasp, 105, 630

\bibitem[{{Saint-Hilaire} \& {Benz}(2005)}]{saint2005}
{Saint-Hilaire}, P., \& {Benz}, A.~O. 2005, \aap, 435, 743

\bibitem[{Smith {et~al.}(2002)Smith, Lin, Turin, Curtis, Primbsch, Campbell,
  Abiad, Schroeder, Cork, Hull, {et~al.}}]{smith2002rhessi}
Smith, D., Lin, R., Turin, P., Curtis, D., Primbsch, J., Campbell, R., Abiad,
  R., Schroeder, P., Cork, C., Hull, E., {et~al.} 2002, Solar Physics, 210, 33

\bibitem[{{Sui} {et~al.}(2002){Sui}, {Holman}, {Dennis}, {Krucker}, {Schwartz},
  \& {Tolbert}}]{sui2002feb20}
{Sui}, L., {Holman}, G.~D., {Dennis}, B.~R., {Krucker}, S., {Schwartz}, R.~A.,
  \& {Tolbert}, K. 2002, \solphys, 210, 245

\bibitem[{{Veronig} {et~al.}(2002){Veronig}, {Vr{\v s}nak}, {Dennis}, {Temmer},
  {Hanslmeier}, \& {Magdaleni{\'c}}}]{veronig2002}
{Veronig}, A., {Vr{\v s}nak}, B., {Dennis}, B.~R., {Temmer}, M., {Hanslmeier},
  A., \& {Magdaleni{\'c}}, J. 2002, \aap, 392, 699

\bibitem[{{Veronig} {et~al.}(2005){Veronig}, Brown, {Dennis}, {Schwartz},
  {Sui}, \& {Tolbert}}]{veronig2005}
{Veronig}, A.~M., Brown, J.~C., {Dennis}, B.~R., {Schwartz}, R.~A., {Sui}, L.,
  \& {Tolbert}, A.~K. 2005, ApJ, 621, 482

\end{thebibliography}

\end{document}